\RequirePackage{snapshot}
\documentclass[]{elsarticle}
\usepackage[utf8]{inputenc}
\usepackage{xcolor}
\usepackage{amsmath}
\usepackage{hyperref}
\usepackage{graphicx}
\usepackage{subcaption}
\usepackage{overpic}
\usepackage{tikz}
\usepackage[normalem]{ulem}

\newcommand{\postPDF}{\ensuremath{F}}

\newcommand{\param}{\ensuremath{\vec{m}}}
\newcommand{\data}{\ensuremath{\vec{d}}}
\newcommand{\cdens}[2]{q\left(#1|#2\right)}
\newcommand{\MCmove}[2]{b\left(#1, #2\right)}
\newcommand{\hastingsR}[2]{r\left(#1,#2\right)}
\newcommand{\lrp}[1]{\left(#1\right)}
\newcommand{\lrb}[1]{\left[#1\right]}
\newcommand{\NormDist}[1]{\ensuremath{\mathcal{N}\lrp{#1}}}
\newcommand{\obj}[1]{S\lrp{#1}}
\newcommand{\fwd}[1]{g\lrp{#1}}
\newcommand{\grad}{\ensuremath{G}}
\newcommand{\datanoise}{\ensuremath{\vec{\epsilon}}}
\newcommand{\ensize}{N}
\newcommand{\todo}[1]{}
\newcommand{\rw}[1]{}

\journal{Journal of Computational Science}
\begin{document}


\begin{frontmatter}

\title{Verification of a real-time ensemble-based method for updating earth model based on GAN}
\author[norce]{Kristian Fossum\corref{mycorrespondingauthor}}
\ead{krfo@norceresearch.no}
\author[norce]{Sergey Alyaev}
\author[norce]{Jan Tveranger}
\author[hw]{Ahmed H. Elsheikh}
%

\address[norce]{NORCE Norwegian Research Centre, Bergen, Norway}
\address[hw]{School of Energy, Geoscience, Infrastructure and Society, Heriot-Watt University, Edinburgh, United Kingdom}
\cortext[mycorrespondingauthor]{Corresponding author}



\begin{abstract}

The complexity of geomodelling workflows is a limiting factor for quantifying and updating uncertainty in real-time during drilling. 
We propose Generative Adversarial Networks (GANs) for parametrization and generation of geomodels, combined with Ensemble Randomized Maximum Likelihood (EnRML) for rapid updating of subsurface uncertainty. 
This real-time ensemble method combined with a highly non-linear model arising from neural-network modeling sequences might produce inaccurate and/or biased posterior solutions. 
This paper illustrates the predictive ability of EnRML on several examples where we assimilate local extra-deep electromagnetic logs. 
Statistical verification with MCMC confirms that the proposed workflow can produce reliable results required for geosteering wells.


\end{abstract}

\end{frontmatter}

\section{Introduction}
\label{sec:introduction}
The process of drilling wells for hydrocarbon production represents a major cost in petroleum reservoir development. 
However, drilling of new wells is necessary to increase the total oil recovery. 
To maximize the value for each drilled well it is necessary to optimize the placement of the well within the reservoir structure. 
An optimally placed well will mobilize more of the petroleum resources, and reduce the need for injected water -- reducing the environmental impact of oil production.

To place a well in its optimal position, operators apply geosteering. 
Here, the well trajectory is adjusted while drilling in response to real-time measurement of the geology surrounding the drill bit. 
The value of geosteering has been well documented in the literature~\cite{Al-Fawwaz2004,Guevara2012,Janwadkar2012}.

The main objective with geosteering is to utilize the information in the measurements to make optimal decisions. 
Hence, geosteering can be seen as a sequential decision process under uncertainty and should be treated in a probabilistic framework \cite{kullawan2014decision}.
Recently, a workflow based on the Ensemble Kalman Filter (EnKF)~\cite{Evensen1994} 
has been employed to condition the geological model on measurements acquired while drilling~\cite{Chen2015b,Luo2015}.
In the EnKF, the uncertainty is represented
by an ensemble of equiprobable realizations.
This workflow has then been combined with a global optimization method and
applied as a Decision Support System (DSS)\cite{Alyaev2019a}.

The DSS framework provides high quality decisions on synthetic cases, and outperforms most of geosteering experts in a controlled experiment \cite{alyaev2021interactive}.
However, practical challenges should be addressed for it to be applicable to real operations~\cite{Alyaev2019a}.
This includes modeling of modern commercial tool to process real measurements as well as real-time earth model that can handle realistic geological complexity.
The forward deep neural network (FDNN) trained on synthetic data for extra-deep electromagnetic measurements \cite{alyaev2021modeling} enabled the real-time ensemble-based update of layered models in 1.5D \cite{jahani2021ensemble}, and 3D \cite{fossum2021reducing}.
Moreover, \cite{rammay2022probabilistic} showed that the model errors present in the FDNN approximation can be alleviated during the ensemble based inversion for the layered case.

\begin{figure}
    \centering
    \includegraphics[width=\textwidth]{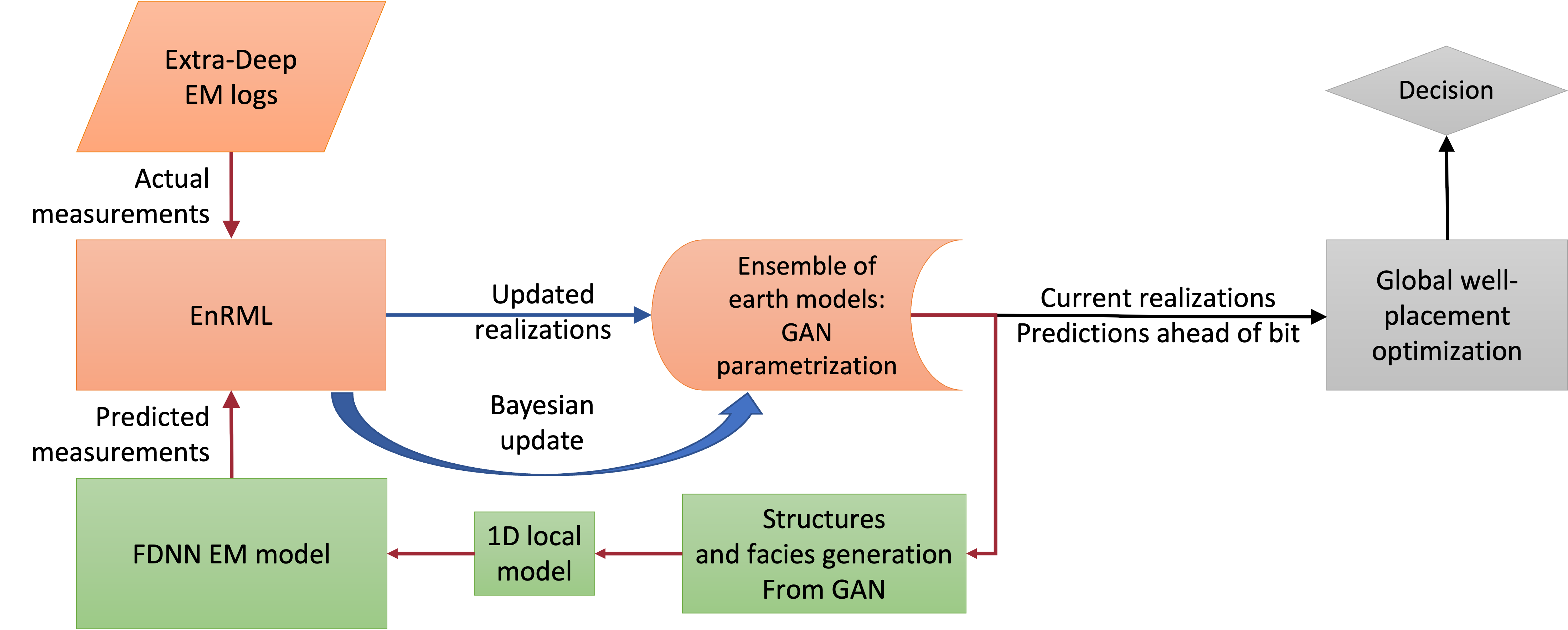}
    \caption{The proposed DSS workflow. 
    Green boxes highlight the new modeling sequence introduced in \cite{fossum2021reducing}. 
    The gray boxes indicate the decision optimization, which is not explored in detail here.}
    \label{fig:DSS_workflow}
\end{figure}

Fossum et. al~\cite{fossum2021bugged} proposed a new modeling sequence which combines the FDNN with a generative adversarial network (GAN) to produce complex geological realizations in real-time to aid geosteering, see Figure \ref{fig:DSS_workflow}.
The premise of the GAN is that it allows to represent the earth model by a Gaussian distribution, where all produced realizations also maintain geological realism. 
This allows using EnKF-like methods to update not only continuous properties but also complex geological structures, which is required for geosteering.
However, the workflow implementation in \cite{fossum2021bugged} converged only with a little starting uncertainty.
\cite{alyaev2021probabilistic} improved the results and demonstrated visually-convincing structural ahead-of-bit prediction on a selected example.
It is known, however, that the approximate ensemble based methods, such as EnKF and its derivatives, can be sensitive to non-linearities present in complex modeling sequences and thus predictions may be biased.


In this paper we present a robust and improved implementation of the framework presented in~\cite{fossum2021bugged} that is able to account for an appropriate starting uncertainty.
Further, we aim to test the convergence properties of the iterative ensemble randomised maximum likelihood (EnRML) method when updating the GAN-based geomodels with FDNN approximation of measurements -- denoted the GAN-FDNN modeling sequence. 
The EnRML probabilistic output is compared to a gold standard Markov Chain Monte-Carlo (MCMC) solution for the same problem using various metrics. The numerical examples demonstrate that the EnRML -- applied to the GAN-FDNN modeling sequence -- generates posterior samples with excellent predictive capabilities that are good approximations to the true posterior solution.  



To construct a reference earth model 
we generate realizations of a fluvial geological environment using a commercial software.
These realizations are then sub-sampled to form a training dataset for the offline training of a Generative Adversarial Network (GAN). 
The GAN is then used, online, to generate plausible geological realizations from a low-dimensional Gaussian input vector. The complete earth modeling loop is described in Section~\ref{sec:GAN}.
For modeling the extra-deep EM measurements we use a forward deep neural network (FDNN) trained on a dataset generated using a commercial simulator~(Section~\ref{sec:proxyLog}).
%
%
%
In Section~\ref{sec:DA} we discuss the exact and the approximate data assimilation (DA) methods. 
The two numerical experiments, designed to test the applicability of our proposed method, are derived, and the numerical results are presented in~Section~\ref{sec:num_exp}.
Finally, we summarize and conclude the paper in Section~\ref{sec:sum_conc}.

\section{Earth modeling using GAN}
\label{sec:GAN}

GANs are a class of unsupervised machine learning methods which can learn to generate new formatted data with the same statistics as the training set.
Motivated by successful applications of GANs for modeling channelized structures for reservoir-simulation workflows \cite{Chan2019a,Chan2019,laloy2021approaching,razak2022conditioning,zhang2021reconstruction}, we use a GAN for efficient earth modeling.

The GAN  consists of two deep neural networks (DNNs): a generator and a discriminator. 
The generator takes a random Gaussian low dimensional vector as input and generates a realization of formatted data: geological realization.
The discriminator takes the formatted data and gives a probability of it being 'real', i.e., belonging to the training set. 
During training the DNNs contest each other in a min-max game.
They are trained simultaneously. 
On each training step the generator creates (fake) geological realizations from the random vectors. 
Fake geological realizations are combined with random samples of the real earth model and are fed to the discriminator.
The loss function for the  generator is  proportional to number of 'fakes' correctly identified by the discriminator.
The loss function for the discriminator is proportional to the total misjudged data samples. 
In our study we use an adapted Wasserstein GAN \cite{Arjovsky2017} with hierarchical deep convolutional networks \cite{radford2015unsupervised} for the generator and the discriminator, see \cite{Arjovsky2017} for implementation details. 

\begin{figure}
    \centering
    \includegraphics[width=0.95\textwidth]{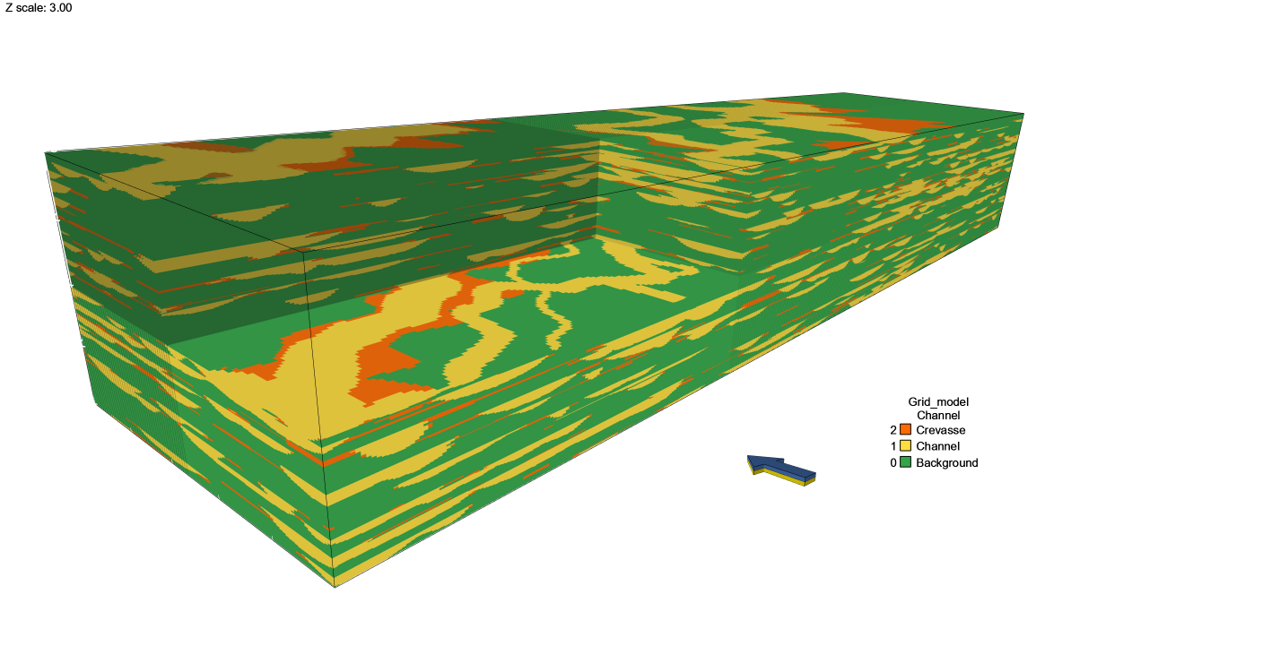}
    \caption{The original earth model generated by the commercial tool. }
    \label{fig:originalModel}
\end{figure}

For geosteering we want to reproduce likely geological realizations of facies and porosity distributions on a 2D vertical geological section along the well to identify the oil-bearing sands ahead of bit.
For training of the GAN we use a large (compared to the area of prediction) reference earth model,
which should provide a realistic test case for the present study in terms of scale and actual geological features and properties.
The reference earth model is constructed using a commercial software that models a synthetic structural framework, a facies model setup derived from outcrop analogue data, and synthetic petrophysical properties of individual facies derived from published literature. 
The resulting model measures 4000m x 1000m x 200m (xyz) with cell
dimensions set to 10x10x0.5~m, yielding a regular  grid of size 400x100x400, see Figure~\ref{fig:originalModel}.

The constructed facies model represents a low net/gross fluvial depositional system. 
It was chosen since it provides complex 3D architectures comprising a limited number of facies, which form contrasting 
geometries, see Figure~\ref{fig:originalModel}.  
Input numbers for statistical generation of facies and geometries are derived from a well-documented outcrop of the Cretaceous lower Williams Fork Formation (Mesa Verde Group) at Coal Canyon, Colorado, USA~\cite{Pranter2014,Pranter2011,Trampush2017}.
%

Key parameters of the facies model setup are listed in Table~\ref{tab:tab1}. The model is not intended as a rendering of the outcrop itself and is consequently simplified compared to descriptions of the original outcrop~\cite{Cole2005,Panjaitan2006,Pranter2009,Pranter2011}.
The model contains three facies: Background/shale, Channels and Crevasse splays. 
The probability distribution of channel width in the model is adapted to include “narrow channel bodies”, and stacking of channels accounts for multi-story channels which comprise more than 80\% of the observed channel bodies. The flow direction of the channel system is set towards $45 \pm 10$ degrees. No trends were used to condition the spatial distribution of channels. The details of this synthetic model are also described in \cite{alyaev2021probabilistic}.

\begin{table}
\centering
\caption{Parameter settings for facies models. 
}
\label{tab:tab1}
\begin{tabular}{|l|l|l|l|}
\hline
\underline{Volumetric fraction} & Value & Tolerance & Comments\\
\hline
Channel system volume fraction &  0.3 & 0.05 & \\
Channel positioning &  1 & & No trends \\
  Crevasse volume fraction & 0.1 & 0.03 & Of channel system vol. frac. \\
  \hline
\end{tabular}
\\
\begin{tabular}{|l|l|l|l|l|}
\hline
\underline{Channel geometry} & Value & SD & Min. & Max.\\
\hline
  Thickness & 4.2 & 1.5 & & \\
  Width & 155 & 50 & 20 & 500 \\
  Correlation W/T & 36 & & & \\
  Amplitude & 400 & 50 & & \\
  Sinuosity & 1.3 & & & \\
  Azimuth & 45 & 10 & & \\
  \hline
\end{tabular}
\\
\begin{tabular}{|l|l|}
\hline
\underline{Form/repulsion} & Setting  \\
\hline
  Cross-section geometry & Parabolic, basic variability \\
  Channel form & Rigid \\
  Repulsion & None \\
  \hline
\end{tabular}
\end{table}

The geological realization is parameterized by a vector of 60 independent parameters.
For each 60-dimensional vector, the generator outputs a 64x64 grid with three values in each grid block.
For a grid block (with dimensions 10.0m along-well and 0.5m thickness) the three values, 'channels',  represent the 
probability of the grid-block belonging to the respective facies class: Background/Channel/Crevasse.
Our generator is also predicting porosity/resistivity distribution within the geo-bodies, but in this study only the facies classes are used.

For training, the original 3D earth model is sampled as 64x64 2D images with three channels. 
The  facies index from the training set is converted into one-hot three-dimensional vector. 
That is, the vector represents the probability of facies: the value of the true index is set to one and other channels to zero.
During evaluation, the resistivity of the facies with the highest probability is applied.




\section{Forward DNN model of extra-deep EM logs}
\label{sec:proxyLog}
To maintain real-time performance of a data assimilation workflow the forward model should be 
fast and support batch, preferably parallel execution. 
Proprietary forward models provided by measurement instrument vendors provide the most accurate results, but they are often not sufficiently fast, and not always optimized for batch execution.
In \cite{alyaev2021modeling}, the authors developed a DNN approximation of such a forward model~\cite{Sviridov2014a}, which we abbreviate FDNN. 

The model approximates the output of the ultra-deep electromagnetic well-bore logging instrument. The instrument is configured to transmit four shallow and nine pairs of deep directional measurements, and has sensitivity to boundaries up to 30 meters to the side from the well bore. We emphasize that the tool provides information around, but not ahead of the drilling position. An illustration of the deep measurements depth of detection is provided in Figure~\ref{fig:joscTruth}.

The input to the FDNN model is a layered geological media with up to three boundaries above and below the measurement instrument as well as the resistivity values of all seven layers. 
In this study we assume that the layer resistivity is isotropic and that the well is aligned with the horizontal axis. 

We produce one synthetic set of measurements for every horizontal position (one per column of cells) of the gridded model which we 'drill' through. 
We choose the most probable facies for each computational cell within the considered column and use the corresponding resistivity value (same as in \cite{alyaev2021probabilistic}): 
\begin{enumerate}
    \item Background, $R = 4.0$ Ohm m;
    \item Channel, $R = 171.0$ Ohm m;
    \item Crevasse, $R = 55.0$ Ohm m.
\end{enumerate}
We find the boundaries between layers composed of pixels with equal resistivities and use the boundaries and the layers' resistivities as the input to the forward model.








\section{Data assimilation during geosteering}
\label{sec:DA}



The DSS for geosteering \cite{Alyaev2019a} uses the data assimilation loop (see Figure \ref{fig:DSS_workflow}, left) to condition the earth model to measurements made while drilling. The fundamental idea is that if a poorly known earth model can be made consistent with measurements in the statistical sense, it will contain non-biased forecasts and, hence, provide a better basis for decisions (see Figure \ref{fig:DSS_workflow}, right).

In this paper, the emphasis is placed on the data assimilation part of the DSS. 
Specifically, we investigate real-time data assimilation
with the EnRML method
utilizing a modeling sequence based on the two neural networks described above:
a GAN-generator for complex earth modeling
and a FDNN models for the synthetic extra-deep EM logs.
The EnRML method is an ensemble-based iterative ensemble smoother which has received a lot of attention for history matching subsurface multi-phase flow problem, see \cite{Oliver2021} and references therein. The method uses an ensemble approximation to the sensitivity matrix, and provides a fast and approximate solution to the Bayesian problem. The method can only be shown to converge for Gaussian posterior distributions. However, the method is known to also sample accurately from moderately non-Gaussian posterior distributions. 
%
To assess the statistical convergence of the EnRML in this context we compare the method to samples generated by a Markov Chain Monte Carlo (MCMC) algorithm. 
Since the MCMC, when properly converged, generates independent and identically distributed samples from the Bayesian distribution, several metrics can be used to assess the statistical error of the EnRML method. The EnRML and the MCMC method is described in detail in the rest of the section.




\subsection{EnRML}
\label{sec:enrml}
The EnRML~\cite{Gu2007} has recently become one of the most successful methods for automatic history matching of petroleum reservoirs. The EnRML is derived by minimization of an objective function using an ensemble approximation of the sensitivity matrix. Since a wide range of methods can be applied for the minimization, the EnRML can be formulated in many different ways. In this study we utilize  the approximate form of the Levenberg-Marquardt method, introduced in~\cite{Chen2013}.

Based on the Bayes' theory, the objective function to be minimized is
\begin{align}
    \obj{\param} = & \frac{1}{2}\lrp{\fwd{\param} - \data_{obs}^*}^TC_{\data}^{-1}\lrp{\fwd{\param}-\data_{obs}^*} \nonumber \\  
    & + \frac{1}{2}\lrp{\param - \param_{prior}}^TC_{\param}^{-1}\lrp{\param - \param_{prior}}.
    \label{eq:obj_func}
\end{align}
Here, $\data_{obs}^*$ is the noisy observed data, $\fwd{\param}$ is the modelling sequence depending on the parameter $\param$, and $\param_{prior}$ is a sample from the prior distribution of the parameters (this is the rough plus smooth sampling approach given in~\cite[chap.10]{Oliver2008}).
Iteration number $i$ of the Levenberg-Marquardt method is given as
\begin{align}
    \delta \param_i =& - \lrb{\lrp{1 + \lambda_i}C^{-1}_{\param} + \grad^T_i C^{-1}_{\data} \grad_i}^{-1} \\
            & \times \lrb{C^{-1}_{\param}\lrp{\param_i - \param_{prior}} + \grad^T_i C_{\data}^{-1} \lrp{\fwd{\param} - \lrp{\data_{obs} + \datanoise}}}
\label{eq:LM_full_update}
\end{align}
where $\lambda_i$ is the Levenberg-Marquardt multiplier, $\grad$ is the sensitivity of data to the parameters, and $\datanoise\sim\NormDist{0,C_{\data}}$ is a realization of the measurement observation noise. 

In the ensemble framework, we approximate $C_{\param}$ and $\grad$ using the ensemble. To this end we define 
\begin{equation}
    \tilde{\grad} = C^{1/2}_{sc}\Delta \data \lrp{\Delta \param}^{-1}
\end{equation}
\begin{equation}
    \tilde{C}_{\param} = \Delta \param \Delta \param^T
\end{equation}
where 
\begin{equation}
    \Delta \param = \lrb{\param_1, \dots, \param_j, \dots, \param_{\ensize}}\lrp{I_{\ensize} - \frac{1}{\ensize}11^T}/\sqrt{\ensize - 1},
\end{equation}
\begin{equation}
    \Delta \data = C_{sc}^{-1/2}\lrb{\fwd{\param_1}, \dots, \fwd{\param_j}, \dots, \fwd{\param_{\ensize}}}\lrp{I_{\ensize} - \frac{1}{\ensize}11^T}/\sqrt{\ensize - 1},
  \end{equation}
$\ensize$ denotes the ensemble size, and $C_{sc}$ is a diagonal matrix for scaling the data, typically containing the measurement variance on the diagonal.
We get the approximate version of the Levenberg-Marquardt update equation by inserting ensemble approximations of $\grad$ and $C_{\param}$,  neglecting the updates from the model mismatch term, substituting the prior precision matrix $C^{-1}_m$ with $\tilde{C}^{-1}_{\param_i}$, and rewriting the equation using the Sherman-Woodbury-Morrison matrix inversion formula~\cite{Golub1983} gives the following update equation
\begin{equation}
  \label{eq:LM_approx_update}
    \delta \param_i = - \tilde{C}_{\param_i}\tilde{\grad}_i^T\lrb{\lrp{1+\lambda_i}C_{\data} +  \tilde{\grad_i}\tilde{C}_{\param_i}\tilde{\grad}^T_i}^{-1}\lrp{\fwd{\param} - \lrp{\data_{obs} + \datanoise}}.
\end{equation}
The update equation is simplified, and made computationally more stable, by inserting the truncated singular value decomposition of $\Delta d$
\begin{equation}
  \Delta \data = U_p S_p V_p^T,
\end{equation}
where the subscript $p$ indicates the number of retained singular values, when they are ordered after descending value. In this work, we define $p$ such that the cumulative sum of the $p$ first singular values equals 99\% of the cumulative sum of all the singular values. Further, we substitute $C_D$ with the ensemble approximation $\tilde{C}_D$
\begin{equation}
  \tilde{C}_{\data}=\Delta \datanoise \Delta \datanoise^T,
\end{equation}
where
\begin{equation}
  \Delta \datanoise = \lrb{\datanoise_1, \dots, \datanoise_j, \dots, \datanoise_{\ensize}}\lrp{I_{\ensize} - \frac{1}{\ensize}11^T}/\sqrt{\ensize - 1}.
\end{equation}
Inserted into~\eqref{eq:LM_approx_update} gives
\begin{align}
  \label{eq:LM_approx_final_one}
  \delta \param_i = -& \Delta \param_i V_p \lrb{\lrp{1 + \lambda_i}S_p^{-1}U_p^TC^{-1/2}_{scl}\Delta \datanoise \Delta \datanoise^TC_{scl}^{-1/2}U_pS_P^{-T} + I} \nonumber \\
                     &\lrp{U_pS^{-1}_p}^TC^{-1/2}_{sc}\lrp{\fwd{\param} - \lrp{\data_{obs} + \datanoise}} \\
                  = -& \Delta \param_i V_p Z \lrb{\lrp{1 + \lambda_i}\zeta + I}^{-1}\lrp{U_pS^{-1}_p Z}^T C^{-1/2}_{sc}\lrp{\fwd{\param} - \lrp{\data_{obs} + \datanoise}},  \nonumber
\end{align}
where $Z$ and $\zeta$ are the eigenvectors and eigenvalues of 
\begin{equation*}
S_p^{-1}U_p^TC^{-1/2}_{scl}\Delta \datanoise \Delta \datanoise^TC_{scl}^{-1/2}U_pS_P^{-T}.    
\end{equation*}

After each application of \eqref{eq:LM_approx_update} we assess convergence and we continue iterations until the scheme is is converged. Here, we consider the method to be converged when the relative difference in the data misfit (the first term in~\eqref{eq:obj_func}) is below a given threshold or when the maximum number of iteration is reached. In the numerical examples, the threshold is $2\times 10^{-2}$ and the maximum number of iterations are $10$. 

\subsection{MCMC}
\label{sec:mcmc}

A reliable method for sampling from a complex posterior distribution is the MCMC technique. MCMC relates to the general framework of methods introduced in~\cite{Metropolis1953} and~\cite{Hastings1970a} for Monte Carlo (MC) integration. 
Firstly, one designs a Markov chain that produce samples from the desired posterior distribution. Secondly, one utilize these samples for MC integration. In this section, the adaptive Metropolis-Hastings method -- the method utilized in the numerical study -- is introduced. For more information on MCMC we refer the reader to~\cite{Brooks2011}, and references therein.

Suppose we want samples from the un-normalized posterior distribution \postPDF, which is the general case with the Bayesian method where the normalizing factor often is very difficult to calculate. 
Assume that the current element of the chain is \param, and one proposes a move to $\param^*$. The proposal is sampled from the proposal distribution $\cdens{\param^*}{\param}$. The move is performed with probability 
\begin{equation}
\MCmove{\param}{\param^*} = min \left(1, \hastingsR{\param}{\param^*} \right)     
\end{equation}
where the Hastings ratio is defined as
\begin{equation}
    \hastingsR{\param}{\param^*} = \frac{\postPDF\lrp{\param^*} \cdens{\param}{\param^*}}{\postPDF\lrp{\param}\cdens{\param^*}{\param}}.
\end{equation}
This is the basis for the Metropolis-Hastings method, and it can be shown that the method generates samples from the posterior distribution $\postPDF$. 

The Metropolis-Hastings algorithm requires a choice of proposal distribution, and some distributions work better than others. The optimal would be to draw proposals directly from the posterior \postPDF. 
However, this is not possible since we cannot sample from this distribution. 
Since the MCMC converges for any proposal distribution that fulfills some general conditions, one idea is to gradually adapt the proposal distribution using previous samples from the chain. 
This adaptive approach ensures a gradually better proposal distribution as the chain evolves. To this end we select the following mixture distribution as our proposal distribution
\begin{equation}
    \param^* \sim \lrp{1-\beta}\NormDist{\param,\lrp{\frac{2.38^2}{N_m}}\tilde{C}_{\param}} + \beta \NormDist{\param, Q_{\param}}.
\end{equation}
Here $\tilde{C}_{\param}$ is the empirical covariance matrix calculated utilizing all the preceding iterations of the Markov Chain, $Q_{\param}$ is some fixed non-singular matrix and $0 < \beta < 1$. Note that $\beta = 1$ until $\tilde{C}_{\param}$ is well defined. Efficient on-line updating of $\tilde{C}_{\param}$ is achieved by the recursion given in~\cite{Dasgupta2007}. This sampling method was applied in~\cite{Fossum2014b,Fossum2015}.
It is well known that the MCMC requires a certain burn-in period since the initial samples are not from the posterior distribution. Hence, it is necessary to monitor the convergence of the method. In this work, convergence is monitored by assessing the maximum root statistic of the multivariate potential scale reduction factor~\cite{Brooks1998}.

\section{Numerical Experiments}
\label{sec:num_exp}

The numerical experiments investigate how the GAN-FDNN modeling sequence can be applied in the data assimilation part of DSS when data assimilation using EnRML is applied for real-time uncertainty reduction.
We design two synthetic experiments that focus on the reduction of uncertainty ahead of measurements and the ability of the algorithm to predict the sand channels in the unexplored part of the geomodel.

To quantify the quality of the EnRML approximation, when applied to the GAN-FDNN modeling sequence, we compare the posterior ensemble of EnRML with true samples from the posterior -- acquired by the MCMC algorithm.
The comparison is done by evaluating several metrics, including visual comparison of standard deviation, and mean, in every point of the domain, point-wise Kolmogorov-Smirnov two-sample test, and visual inspection of kernel density estimates of the marginal distribution of GAN-input vector $m_i$.

\begin{figure}
1. \includegraphics[width=0.45\textwidth]{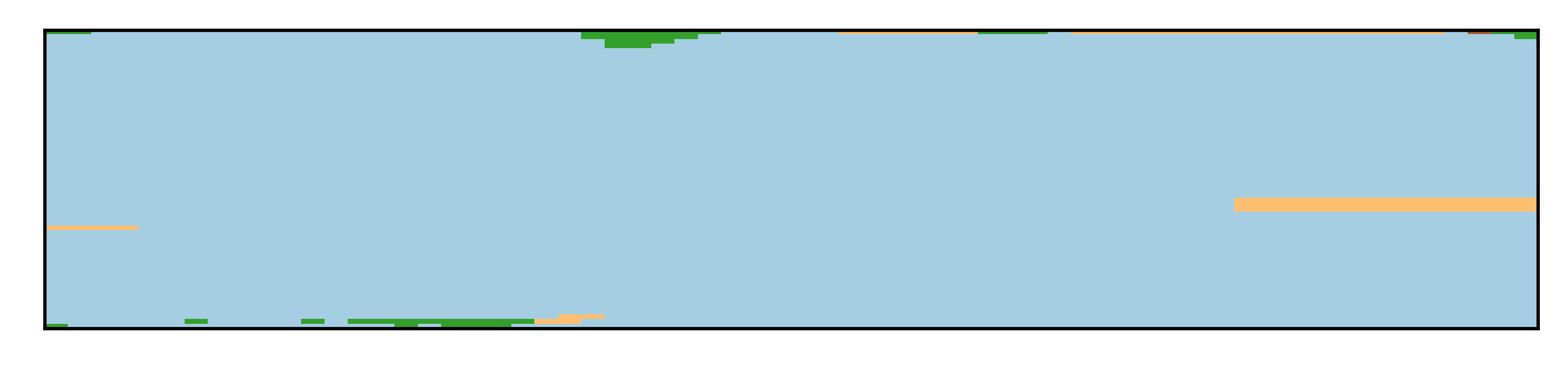}
\includegraphics[width=0.45\textwidth]{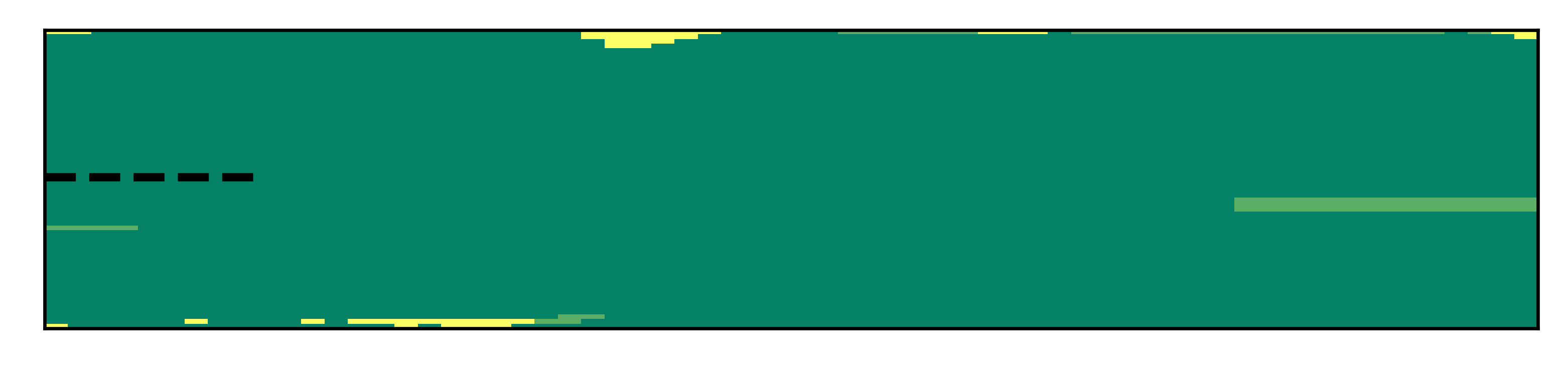}
\includegraphics[trim={18cm 0.2cm 0 0.2cm},clip,width=0.04\textwidth]{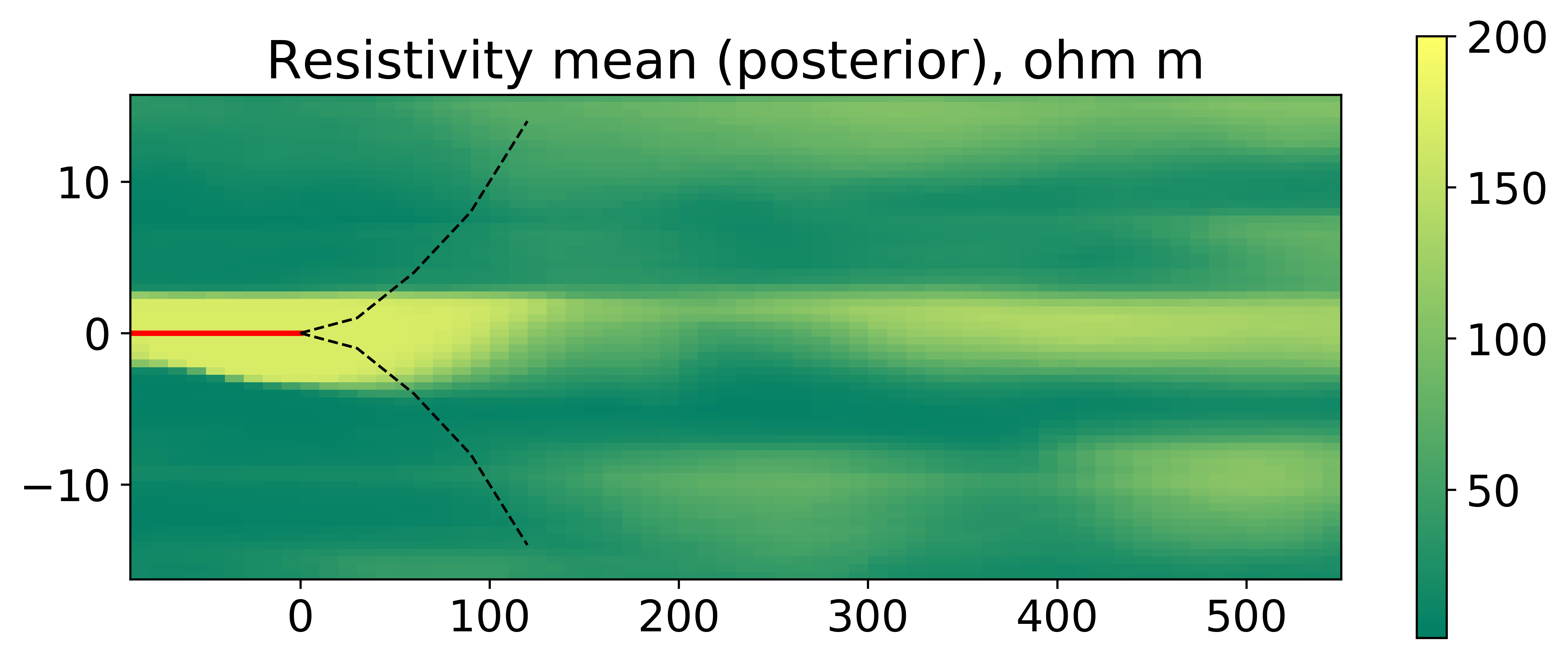}
\\
2. \includegraphics[width=0.45\textwidth]{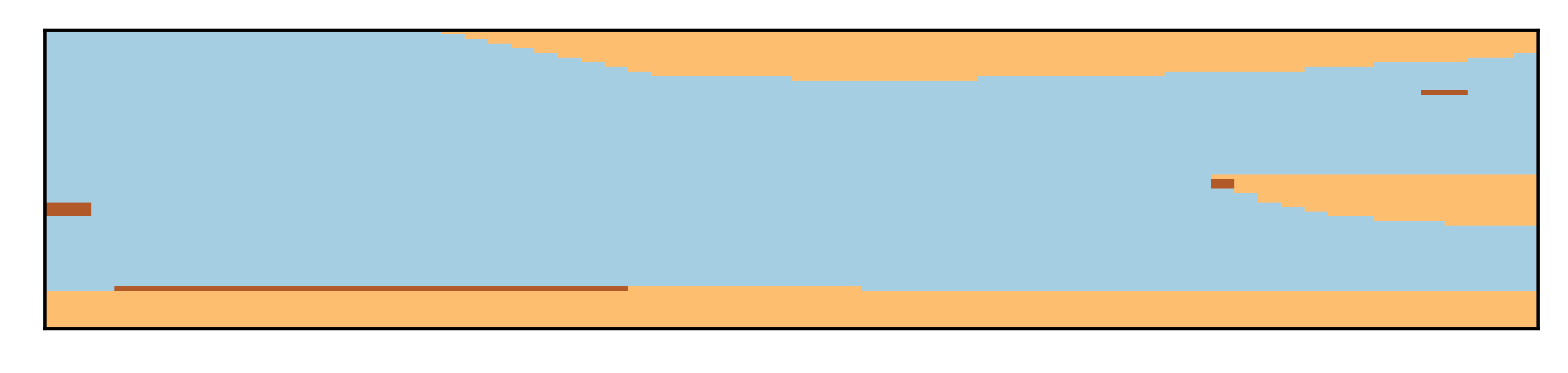}
\includegraphics[width=0.45\textwidth]{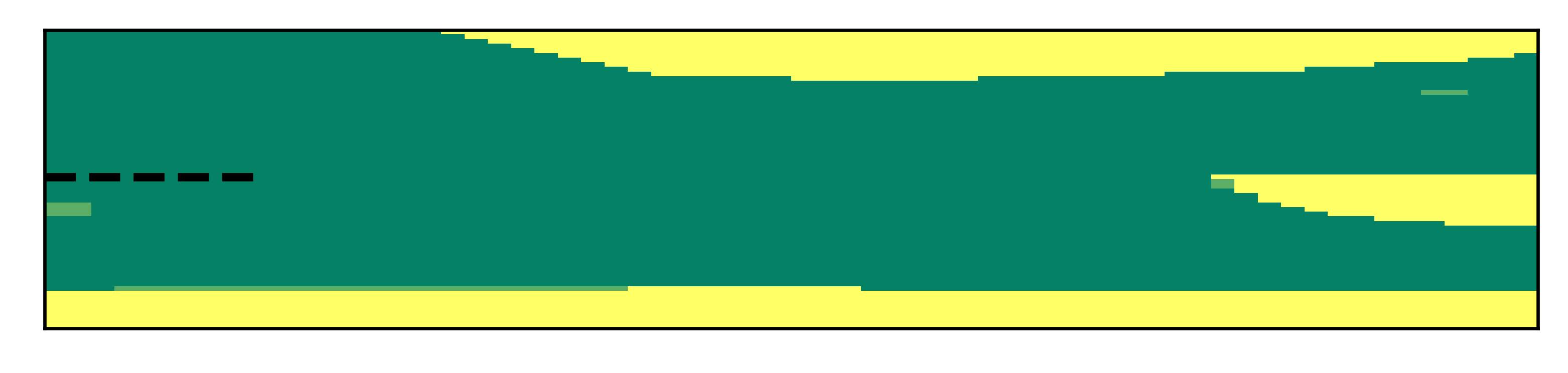}\\
3. \includegraphics[width=0.45\textwidth]{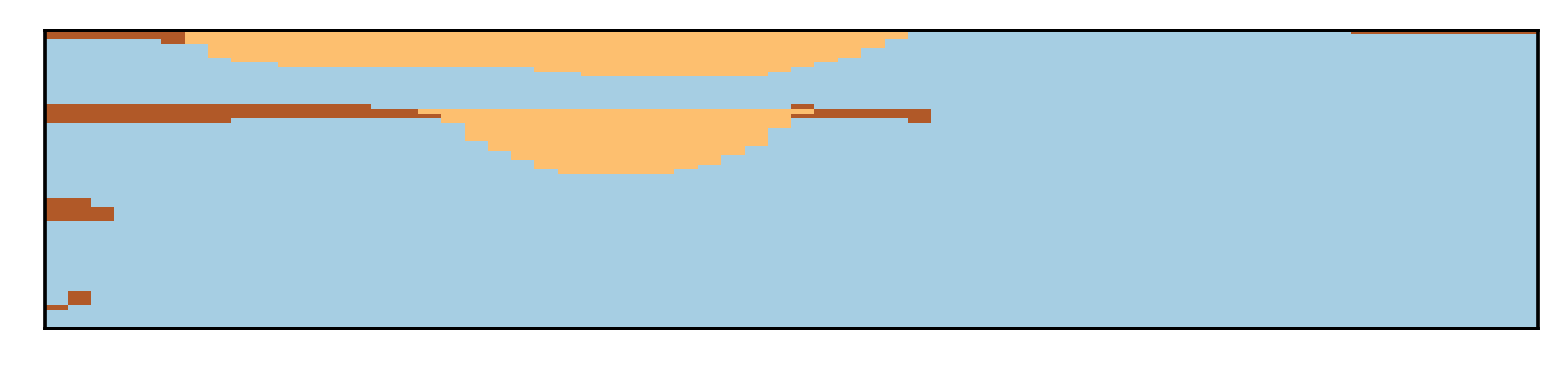}
\includegraphics[width=0.45\textwidth]{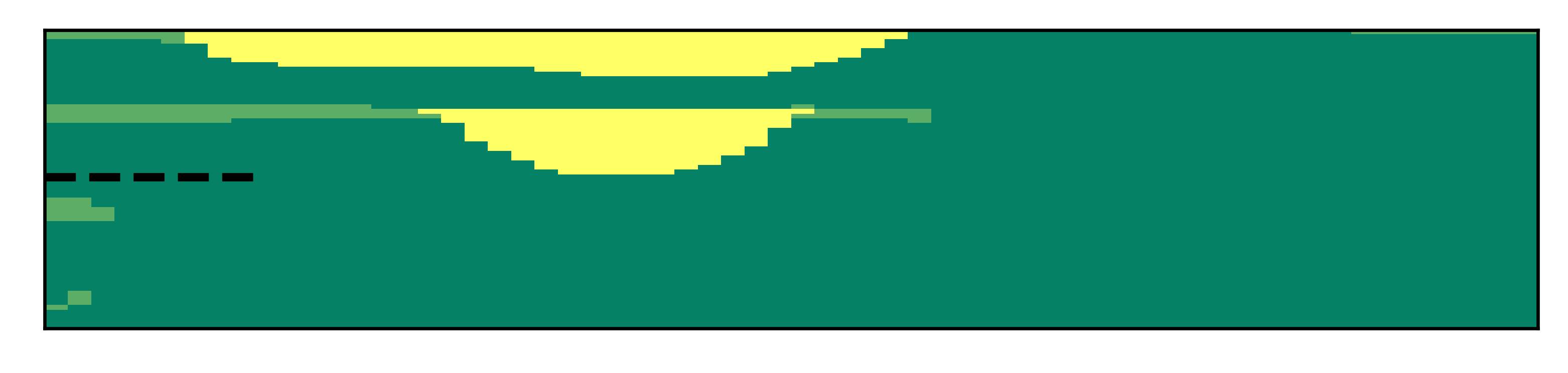}
\\
4. \includegraphics[width=0.45\textwidth]{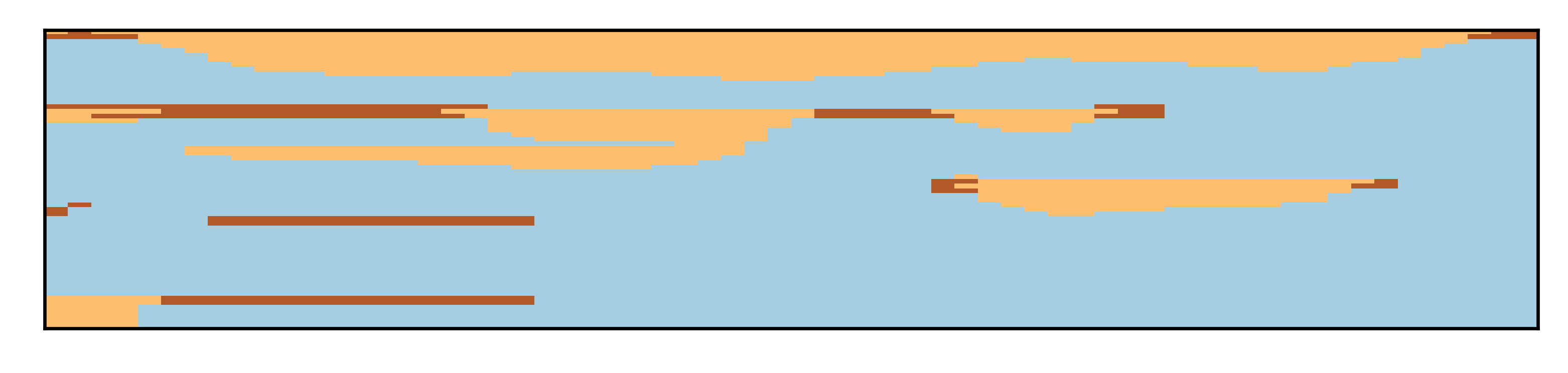}
\includegraphics[width=0.45\textwidth]{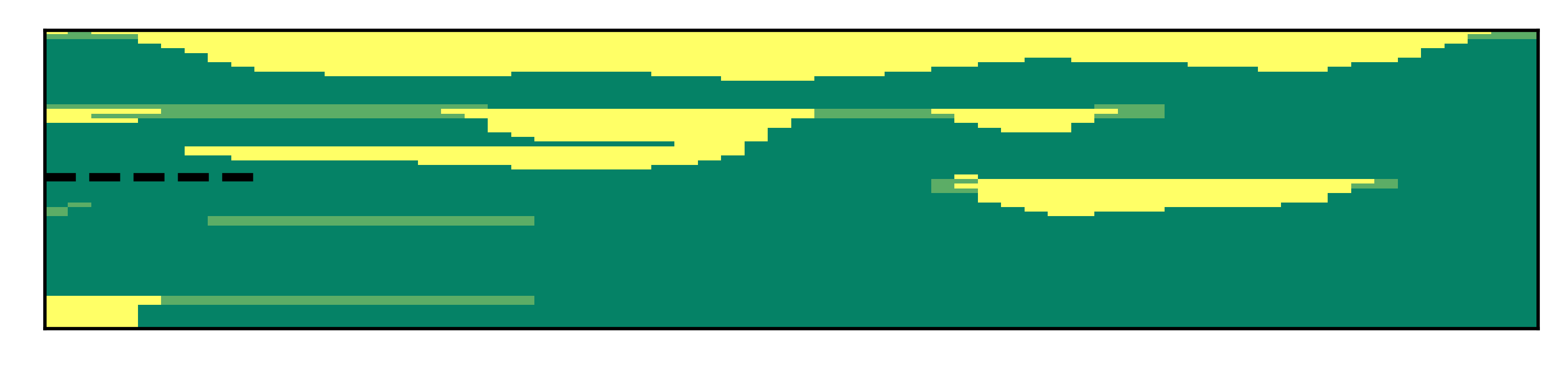}\\
5. \includegraphics[width=0.45\textwidth]{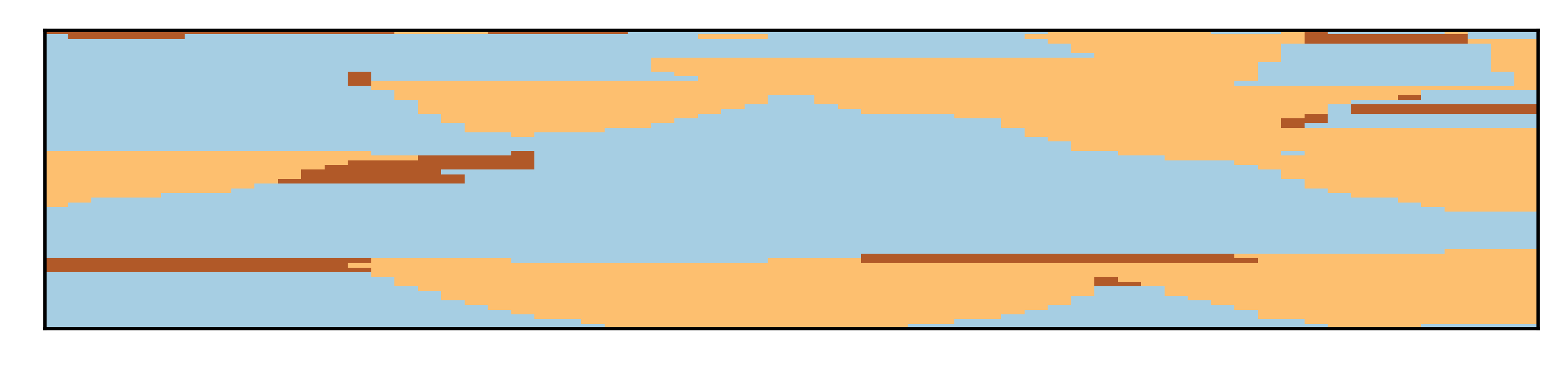}
\includegraphics[width=0.45\textwidth]{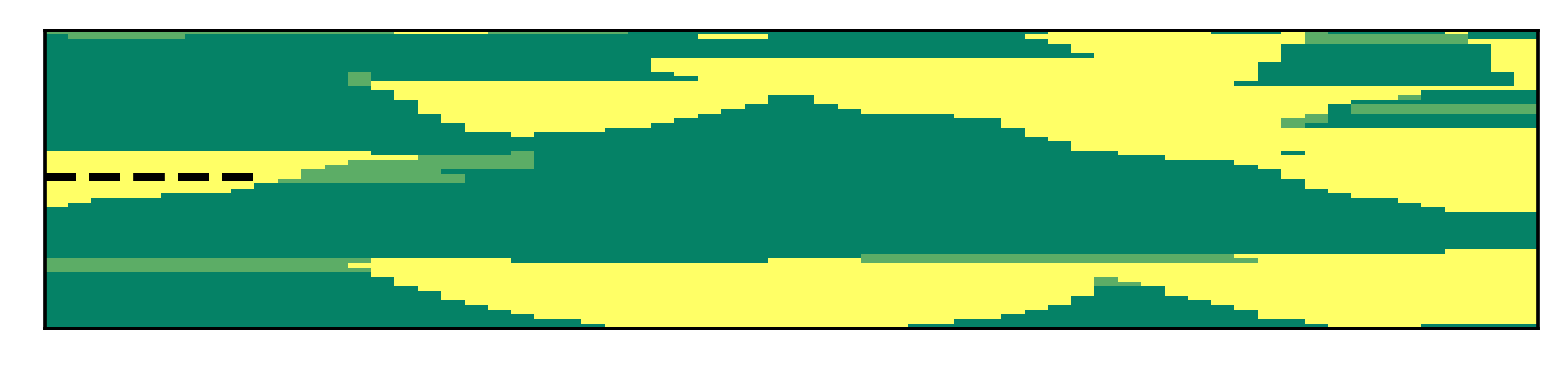}\\
6. \includegraphics[width=0.45\textwidth]{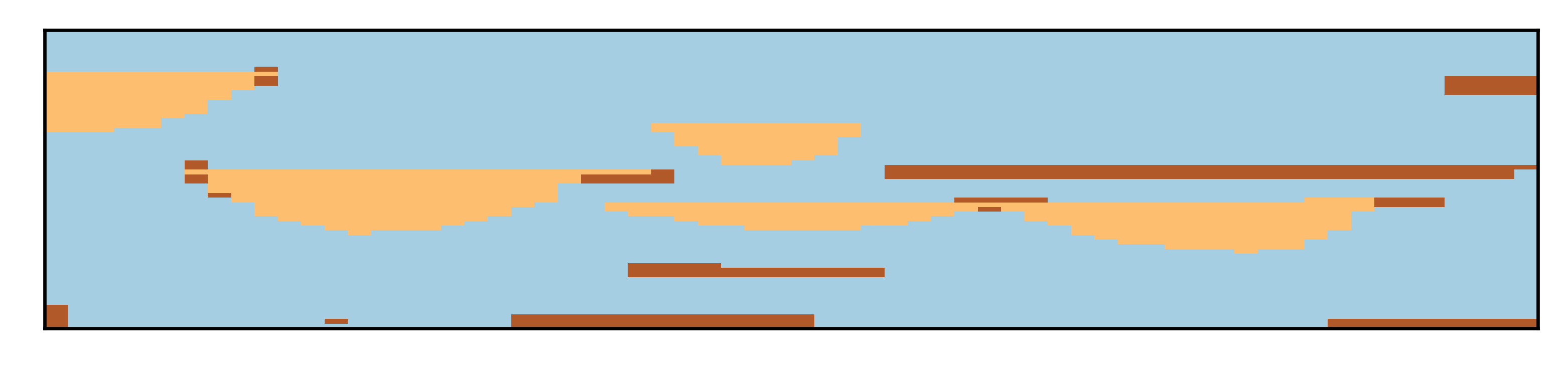}
\includegraphics[width=0.45\textwidth]{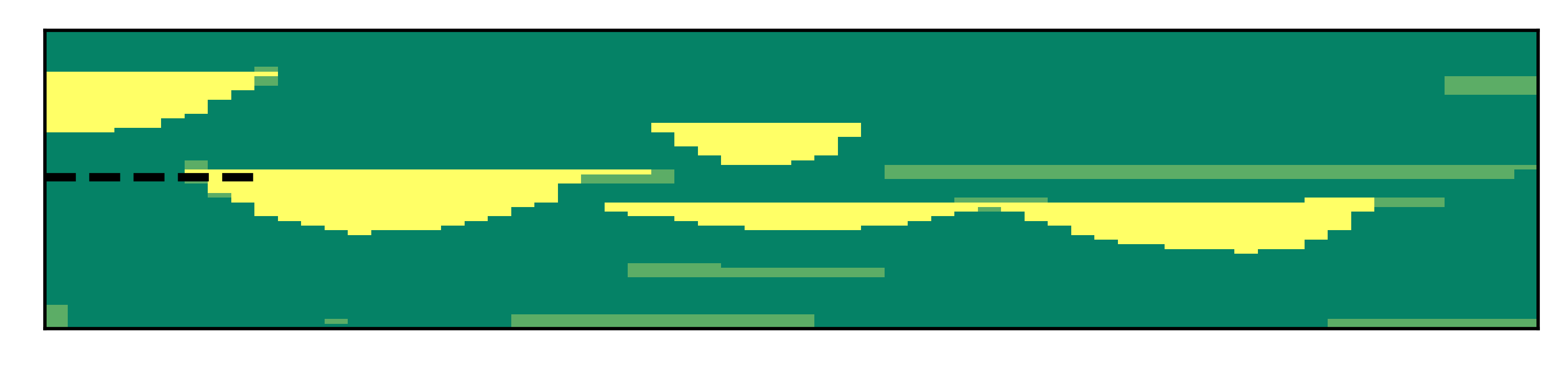}\\
\caption{The rows show the six first realizations from the prior ensemble. The left row shows the facies model (Background/Channel/Crevasse), while the right column shows the derived resistivity image. 
The dotted lines indicate measurement positions.}
  \label{fig:Prior_models}
\end{figure}
\begin{figure}
    \centering
    \includegraphics[trim=0 0 2.5cm 0, clip,height=0.22\textwidth]{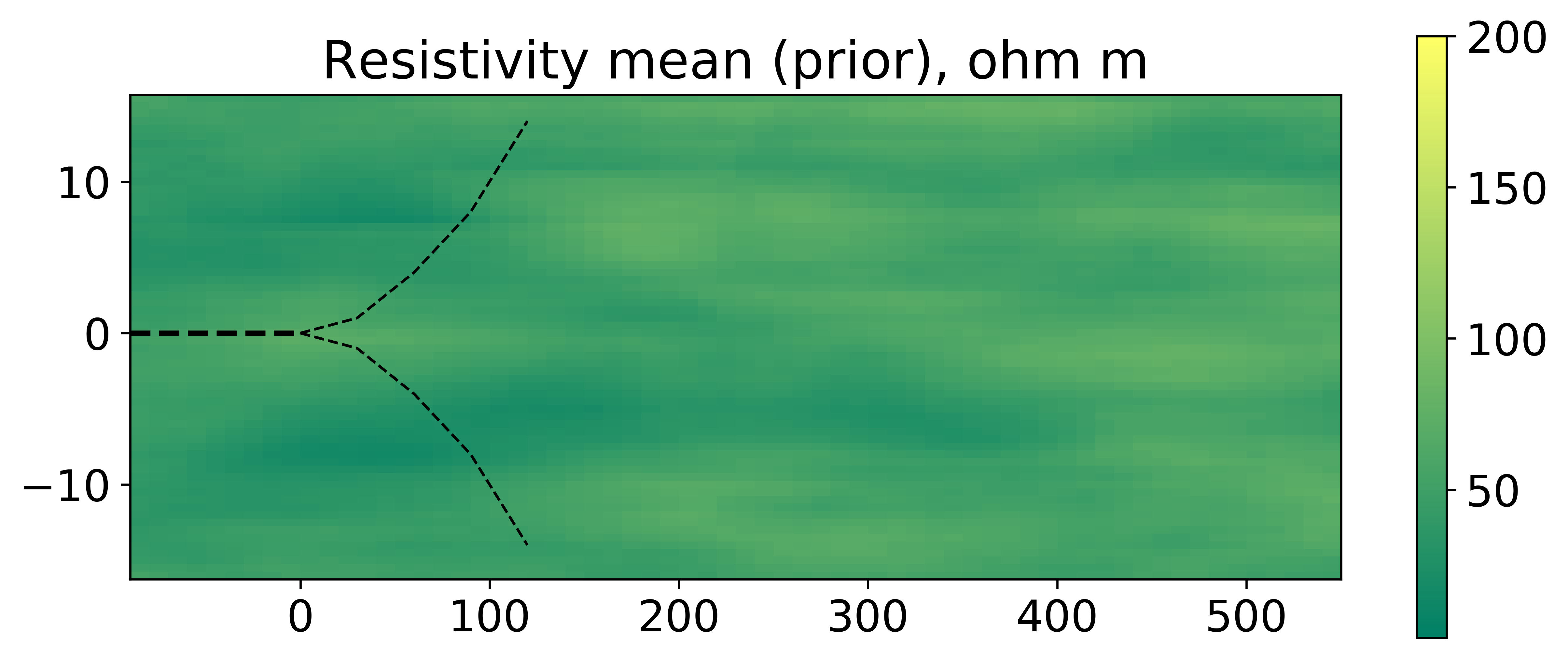}
    \includegraphics[trim=0 0 2.5cm 0, clip,height=0.22\textwidth]{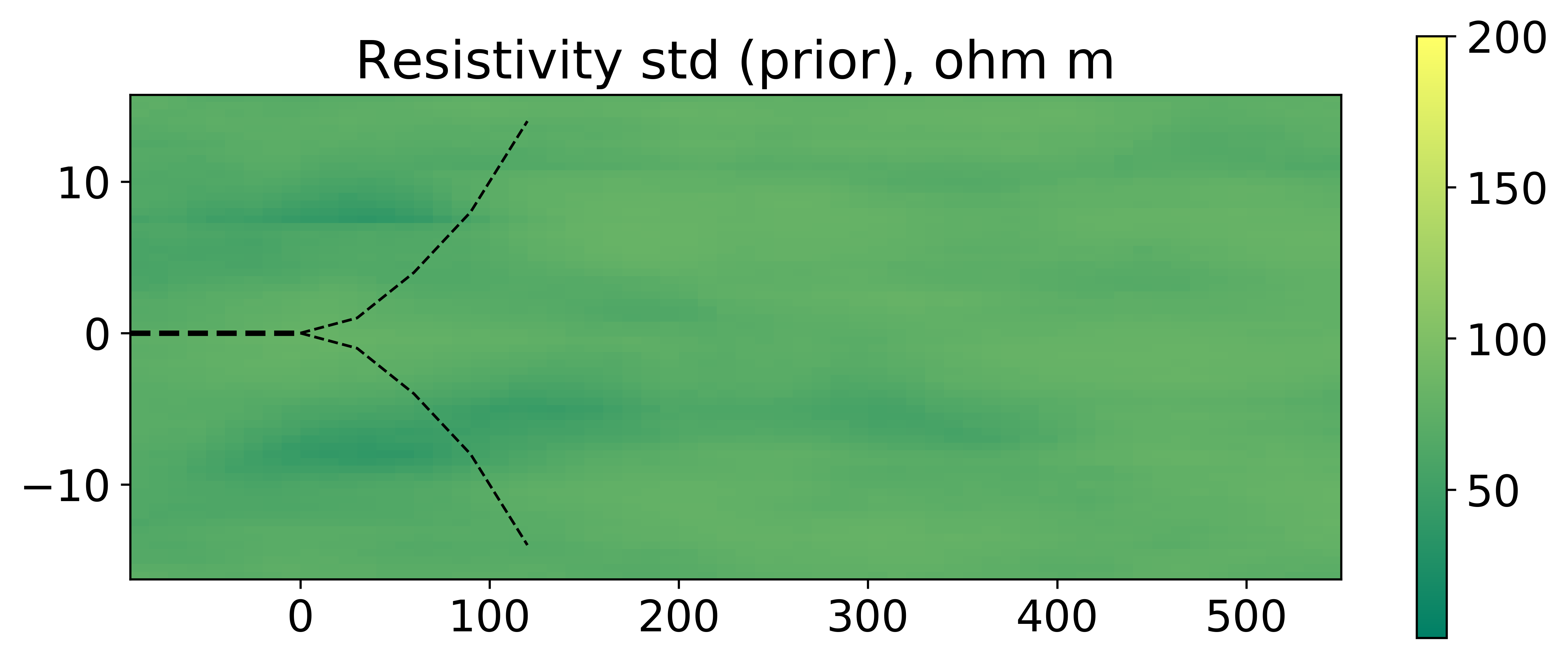}
    \caption{Mean and standard deviation in resistivity models derived from the prior.}
    \label{fig:resitivityPrior}
\end{figure}

We perform two numerical tests. In both tests we utilize the generative neural network, introduced in Section~\ref{sec:GAN}, to represent the earth model with uncertainty. Hence, our goal is to condition the poorly known 60-dimensional input vector, $\param$, to measurements.
The prior realizations of the earth model are generated by applying the generative network to parameters sampled from
a multivariate Gaussian distribution, $\param\sim\NormDist{\hat{\param_0},C_{\param}}$.
The distribution is slightly shifted to simulate conditioning on pre-drill information. 
The $\hat{\param_0}$ represents the shifted mean and is defined by the equation equation:
\begin{equation}
    \hat{m_0}_{i} = \left\{
        \begin{array}{ll}
            0,\quad & i=[20..44],\\
            0.25 {m_0}_{i},\quad & \textrm{otherwise},
        \end{array}
    \right.
\end{equation}
where $\param_0$ is the synthetic truth from \cite{alyaev2021probabilistic}, see Figure \ref{fig:eageTruth}. We use uncorrelated covariance matrix with marginal variance of $C_{m_{i}} = 1$ for all parameters, similar to the GAN training.
Figure~\ref{fig:Prior_models} shows six generated earth model realizations from the prior model. 
From the figure, we observe that this setup provides significant variation in the earth model, which is reflected in the relatively flat mean and the standard deviation derived from the full ensemble of 500 realizations, see Figure \ref{fig:resitivityPrior}.
At the same time, all the realizations are consistent with the chosen channelized geological setting.
We emphasize, that we use the same prior for both numerical experiments.

\begin{figure}
    \centering
    \includegraphics[width=0.7\textwidth]{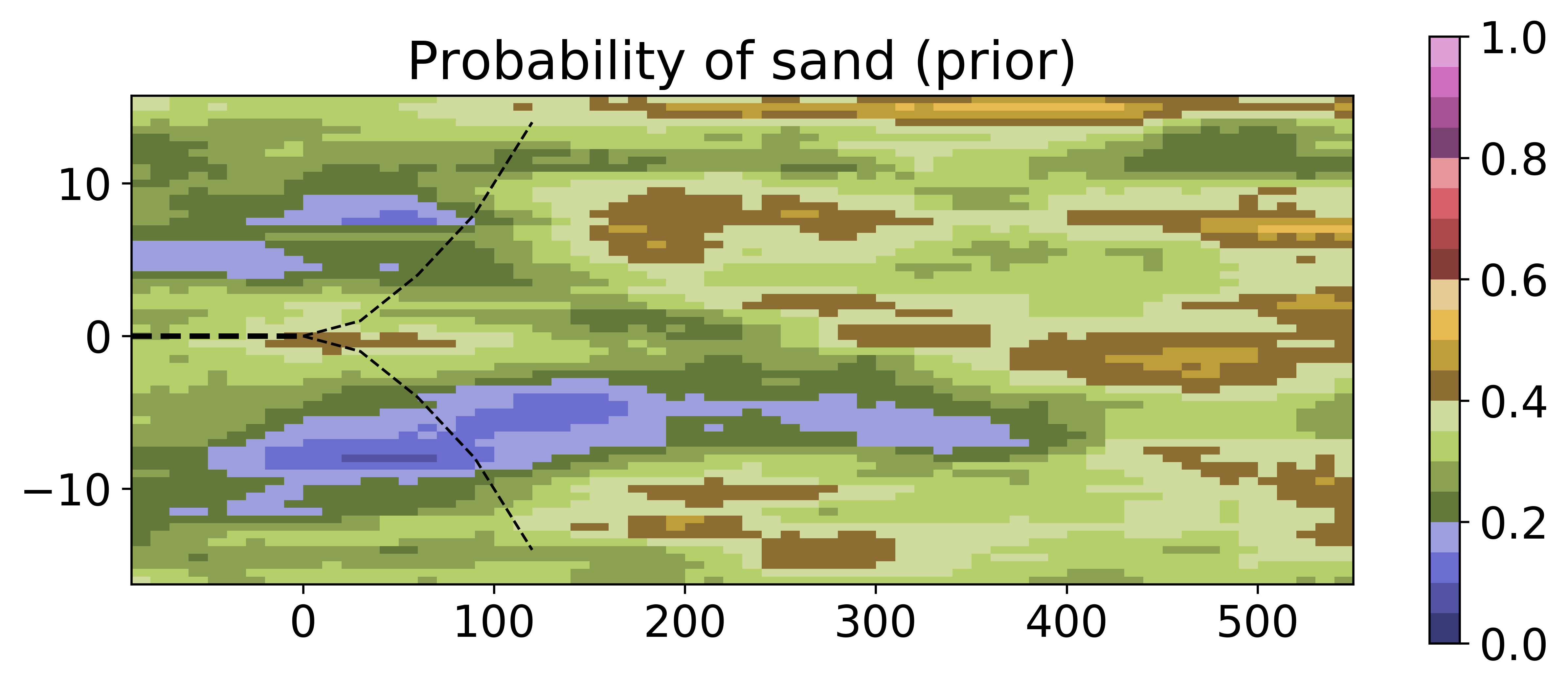}
    \caption{Spatial distribution of the probability of sand facies (crevasse and channel) in the prior model.}
    \label{fig:probabilitySandPrior}
\end{figure}

We conduct two numerical experiments, that differ with respect to the synthetic truth. In the first experiment, the truth model from~\cite{alyaev2021probabilistic} is applied. In the second experiment, the synthetic truth depicted in Figure \ref{fig:joscTruth} is applied. Hence, for experiment 2, the prior is biased towards a wrong model (indicating erroneous pre-drill information) making the data assimilation problem harder.

The numerical study is performed in the same manner for both experiments. Firstly, we sample the true posterior with the MCMC. Here, 8 Markov chains, starting from different initial points, were run for $10^{6}$ steps. At that point, based on assessing the multivariate potential scale reduction factor, the MCMC was found to be converged. Samples from the posterior were then extracted by removing the burn-in phase, and by thinning. For each of the 8 chains, the first half of the chain was removed, and every 100th iteration from the second half of the chain was retained, leaving $4\times 10^4$ samples from the posterior distribution.
Secondly, we estimate the posterior distribution using the EnRML method introduced in Section~\ref{sec:enrml}. Due to the fast simulation time, we utilized an ensemble size of $\ensize=500$, and in addition, we applied the correlation-based localization technique introduced in~\cite{luo2018b}.
Finally, we assess the result from the EnRML by comparison with the samples from the MCMC.

\subsection{Example 1 -- Verification of convergence on an example from literature}

\begin{figure}
    \centering
    \includegraphics[width=0.7\textwidth]{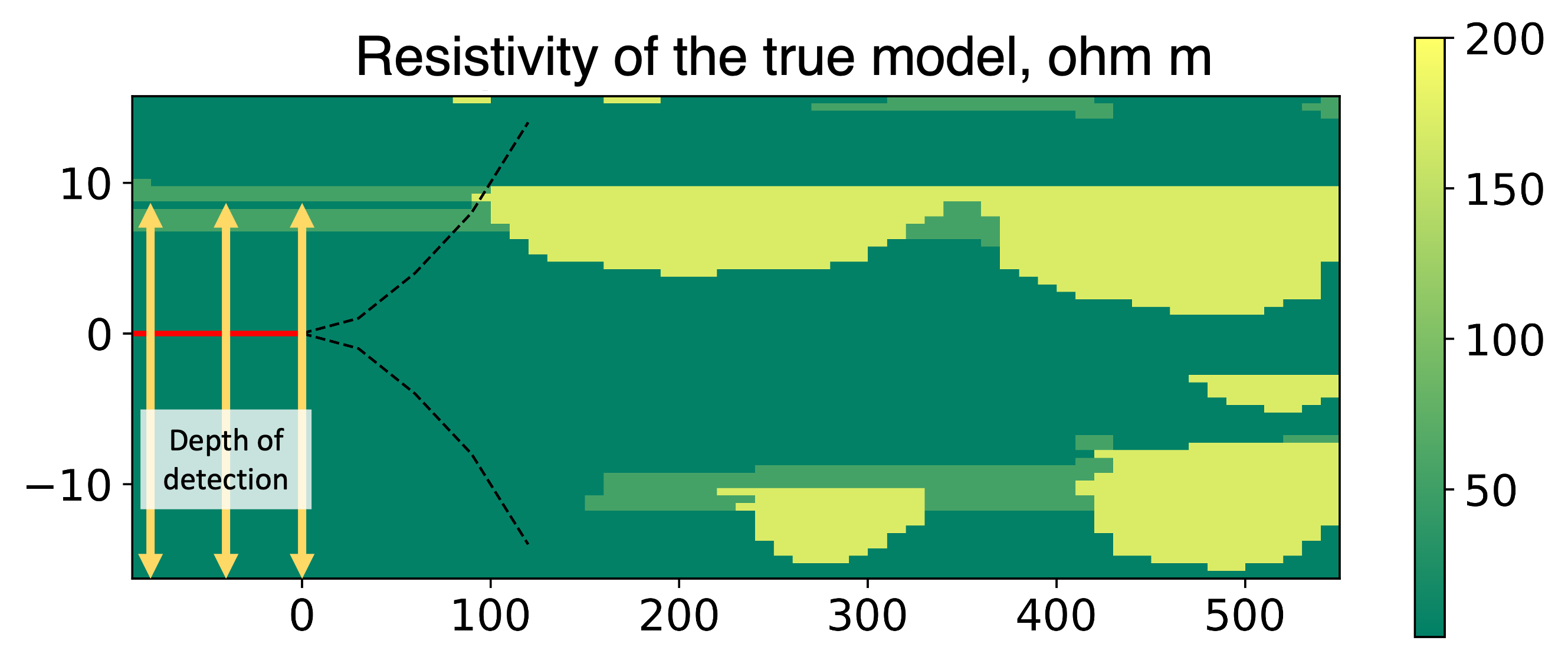}
    \caption{The resistivity of an earth model generated by GAN used as the synthetic truth for Example 1 (adapted from the numerical example in \cite{alyaev2021probabilistic}). 
    The yellow arrows show the region with measurements and their extent illustrates the maximum sensitivity range, termed depth of detection. 
    The filled red line is the drilled well, and the dashed lines indicate the potential for geosteering.}
    \label{fig:eageTruth}
\end{figure}

The first numerical example tests the sampling capabilities for the EnRML on an example from the literature. The synthetic true log is generated from the true model~\ref{fig:eageTruth}. Hence for this case, the prior mean is slightly shifted towards the true model.

 \begin{figure}
 \hspace{0.9cm} {\bf{EnRML}} \hspace{4.5cm} {\bf{MCMC}}\\
    \centering
    \includegraphics[trim=0 0 2.5cm 0, clip,height=0.22\textwidth]{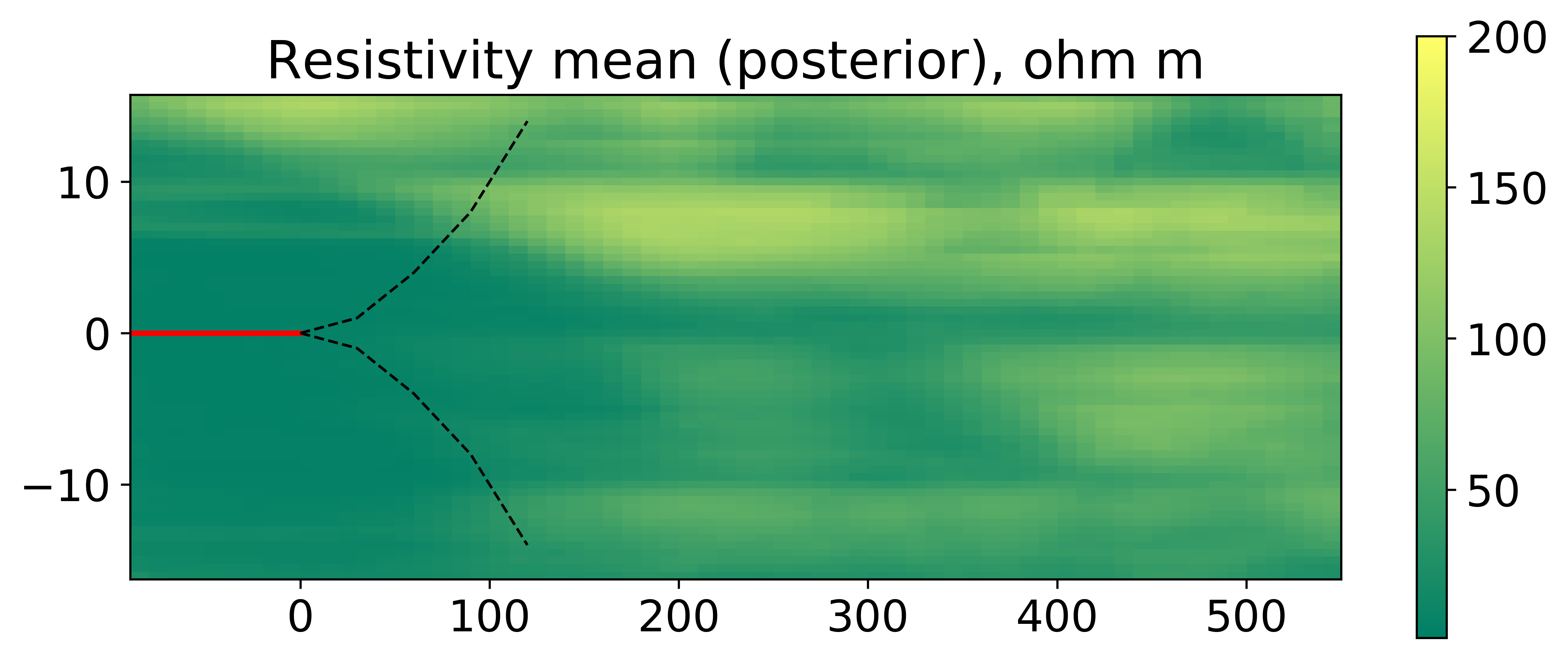}
    \includegraphics[height=0.22\textwidth]{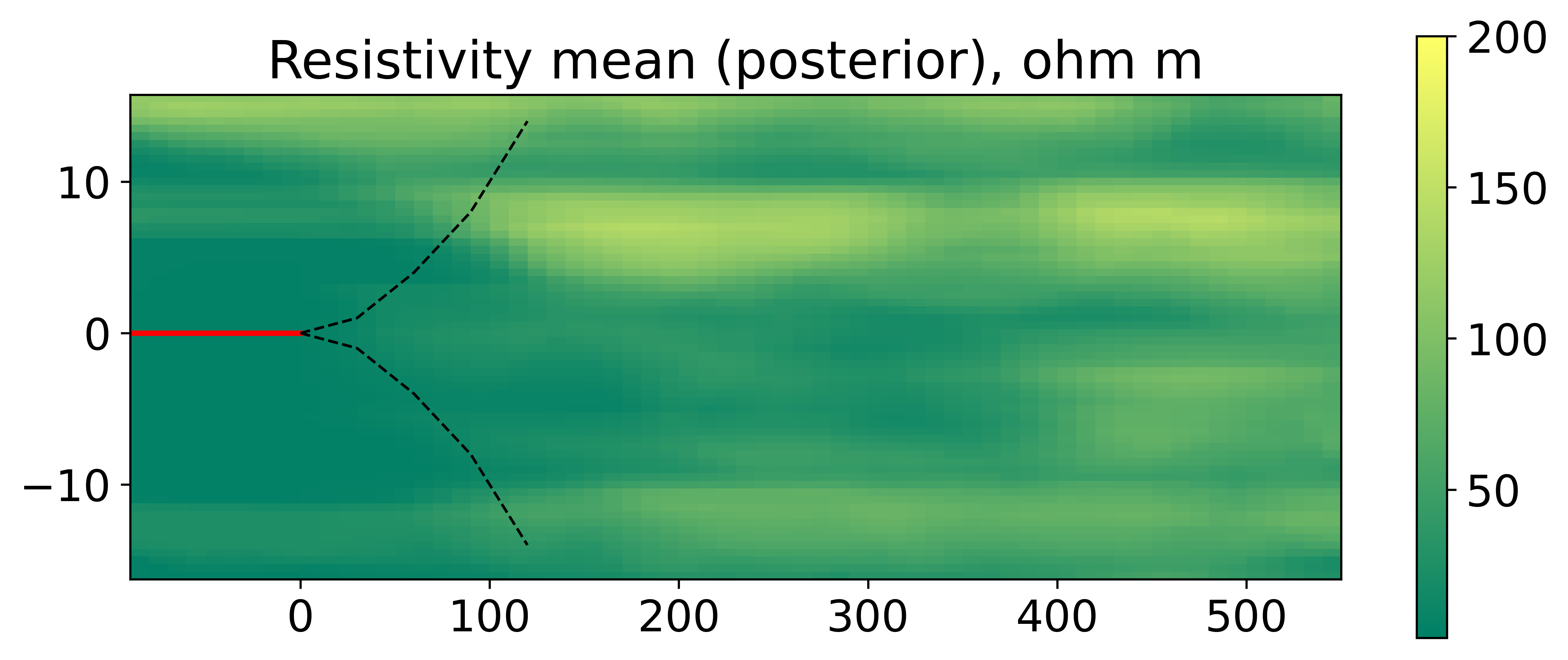}
    \\
    \includegraphics[trim=0 0 2.5cm 0, clip,height=0.22\textwidth]{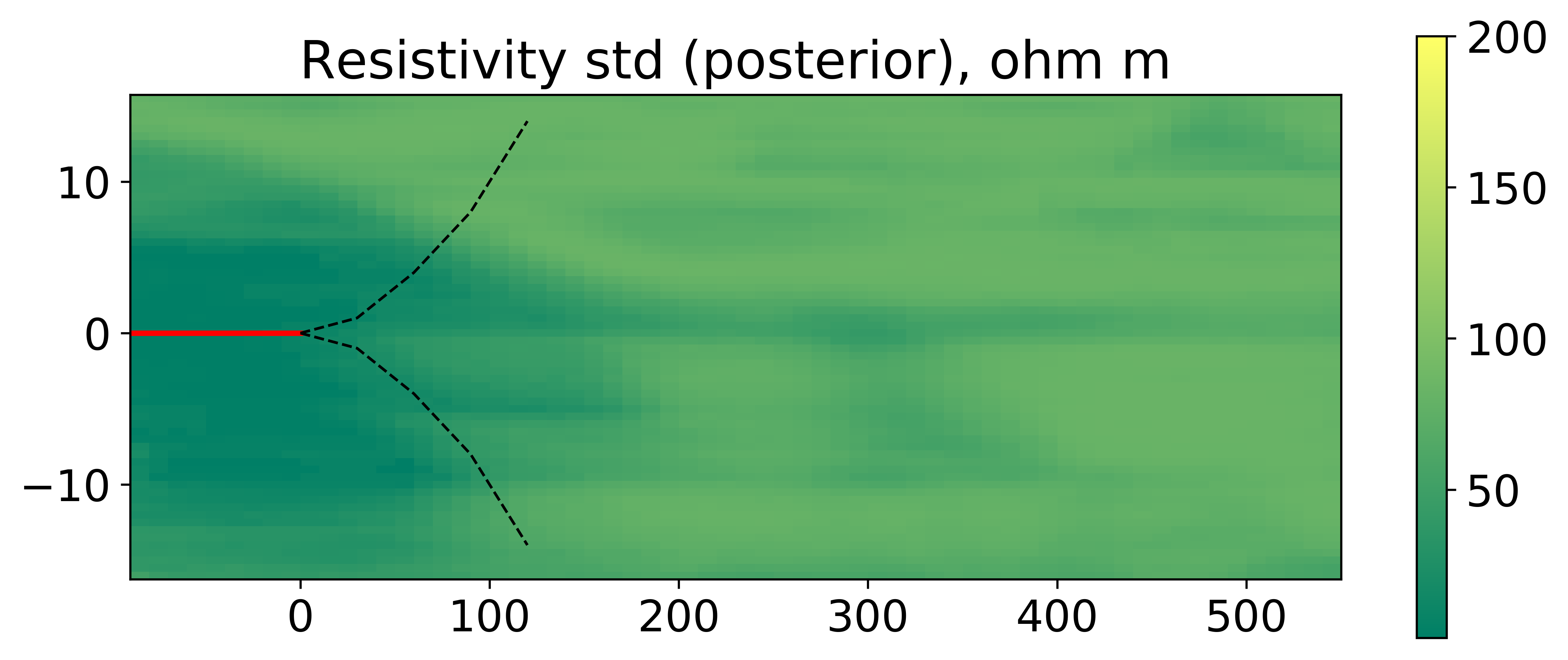}
    \includegraphics[height=0.22\textwidth]{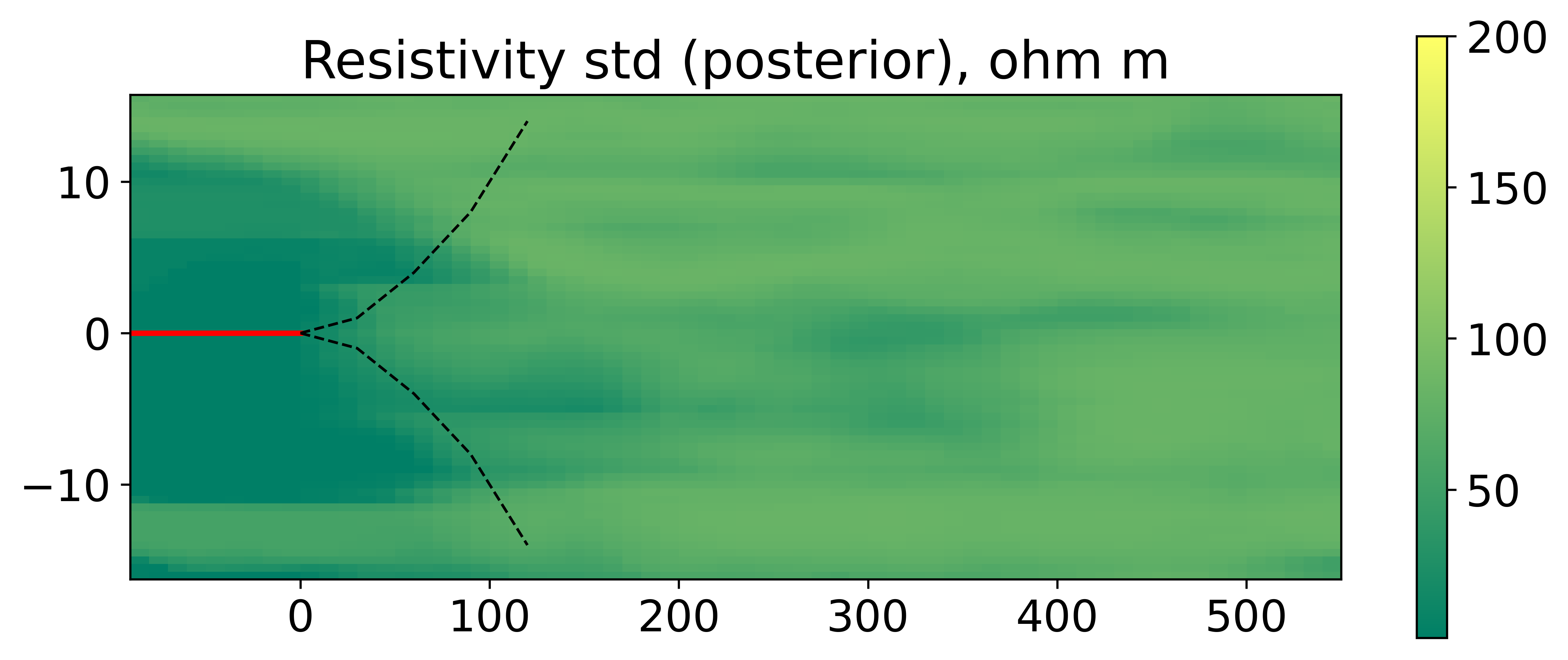}
    \caption{Mean (top) and standard deviation (bottom) in resistivity models derived from the posterior ensembles of EnRML (left column) and MCMC (right column).}
    \label{fig:resitivityPosteriorBoth-ex1}
\end{figure}
Figure~\ref{fig:resitivityPosteriorBoth-ex1} shows the mean and standard deviation of the posterior resistivity model. Compared to the prior mean and standard deviation (shown in figure~\ref{fig:resitivityPrior}) it is clear that the uncertainty around the well is significantly reduced. Moreover, the same reduction is observed for both the EnRML and the MCMC. Apart from slightly sharper boundaries for the MCMC, the EnRML approximation to the posterior mean and standard deviation is almost indistinguishable from the true posterior mean and standard deviation.

 \begin{figure}
 \hspace{0.9cm} {\bf{EnRML}} \hspace{4.5cm} {\bf{MCMC}}\\
    \centering
    \includegraphics[trim=0 0 2.5cm 0, clip,height=0.22\textwidth]{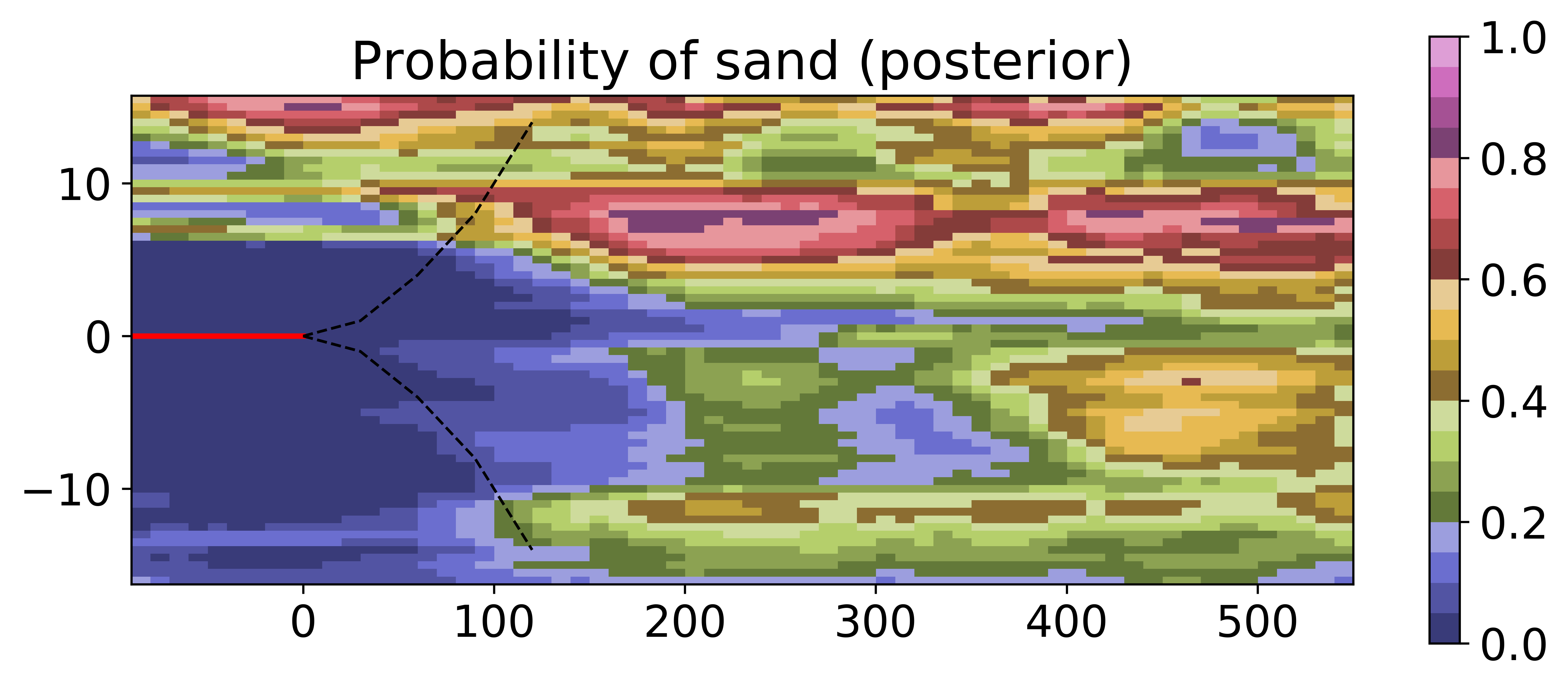}
    \includegraphics[height=0.22\textwidth]{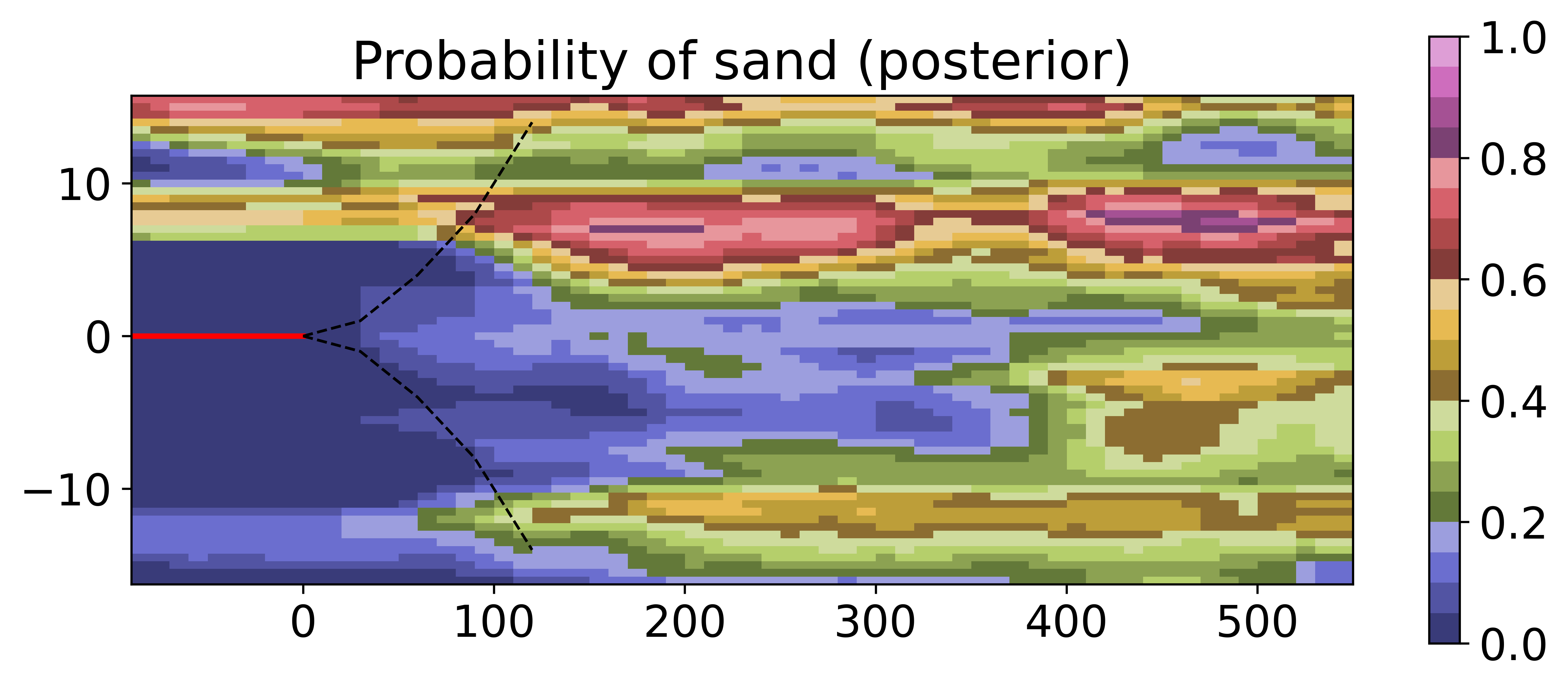}
    \caption{Probability of sand facies (crevasse or channel) from the posterior ensembles of EnRML (left column) and MCMC (right column).}
    \label{fig:prob_good_sand-ex1}
\end{figure}
A similar conclusion can be drawn from the plots showing the estimated point-wise probability of the sand facies, plotted in Figure~\ref{fig:prob_good_sand-ex1}. Around the well, the EnRML and the MCMC are almost indistinguishable. Ahead of the bit, there are small differences. However, the predictive capability, as described in~\cite{alyaev2021probabilistic}, is also present in the MCMC solution. Hence, this is a true feature of the posterior solution with the GAN-FDNN modeling sequence.  

\begin{figure}
    \centering
    \includegraphics[width=0.7\textwidth]{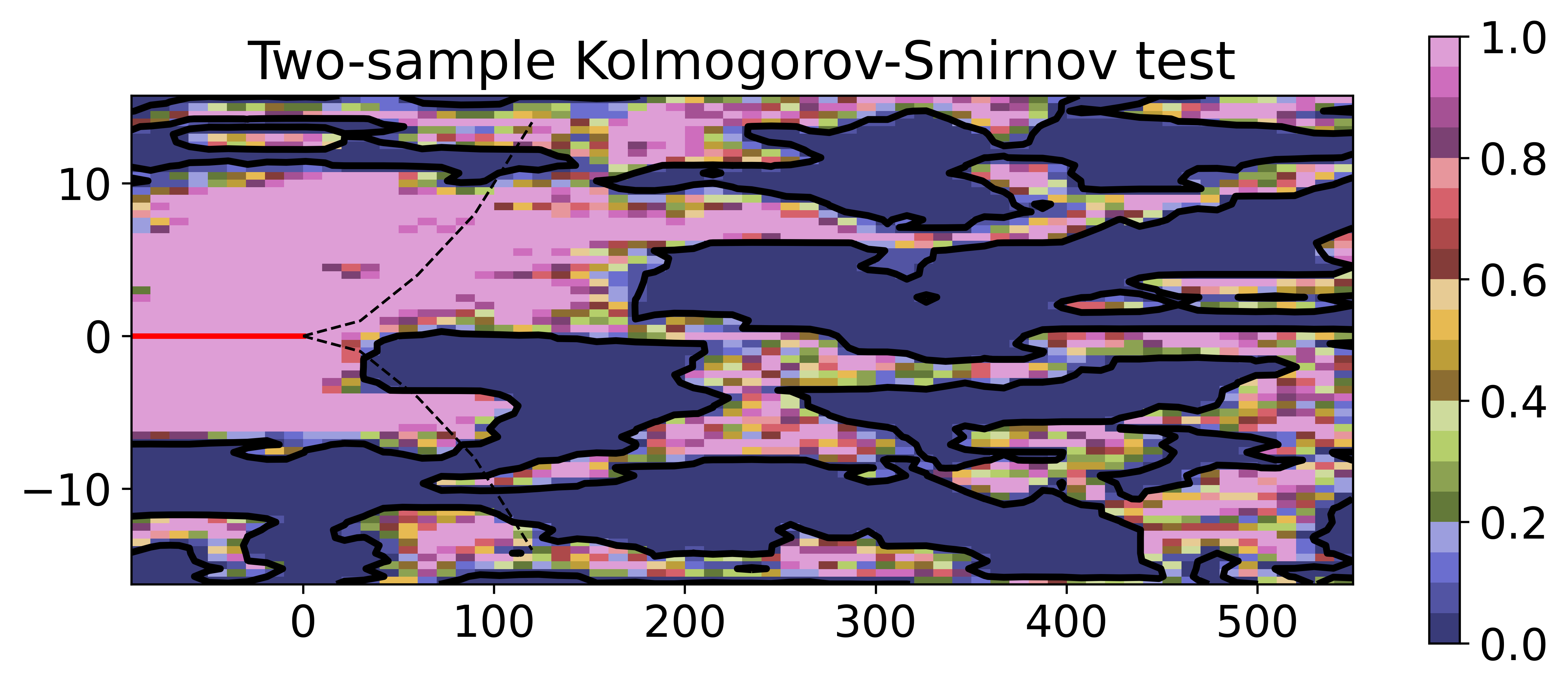}
    \caption{P-value from the Kolmogorov-Smirnov two-sample test of sand facies (crevasse or channel). The Contour line gives the significance level of 0.05.}
    \label{fig:KS_test-ex1}
\end{figure}
To evaluate the statistical distance between the samples from the EnRML and the samples from MCMC we perform a Kolmogorov-Smirnov two-sample test. This is a non-parametric test of equality for one-dimensional probability distributions. The earth model is a 2D image, and not one-dimensional. Hence, we perform the test on the marginal distribution for each cell. The P-values from the test of the H0 hypothesis of equal distributions are shown in Figure~\ref{fig:KS_test-ex1}. The significance level of 0.05 is given by the black contour line. Hence, for all p-values higher than 0.05 one cannot differentiate between the marginal distributions and the H0 hypothesis hold. 

\begin{figure}
    \centering
    \includegraphics[width=0.8\textwidth]{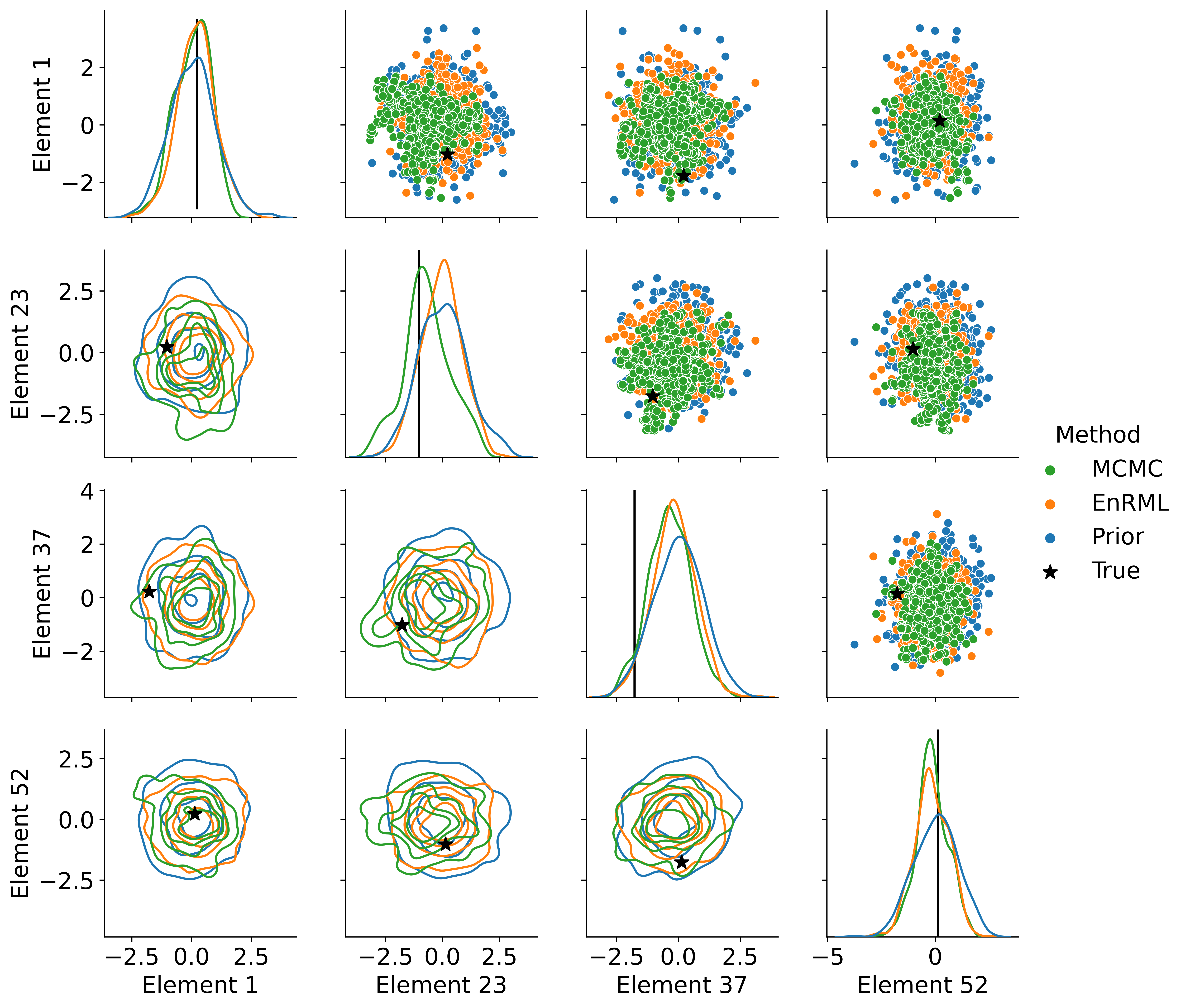}
    \caption{Marginal and bi-variate statistics of selected elements from $\param$.}
    \label{fig:Seaborn-ex1}
\end{figure}
As a final evaluation of the results, we plot a selection of marginal and bi-variate elements of the input vector $\param$. To highlight the effect of the pre-drill information, we selected two elements that was shifted ($m_{1}$ and $m_{52}$) and two that were not shifted ($m_{23}$ and $m_{37}$). In Figure~\ref{fig:Seaborn-ex1} the kernel density estimate of the selected elements is plotted along the diagonal, the scatter plot of the pairwise elements is given in the top corner, while the contours of the 2D Kernel density estimate of the pairwise elements are given in the lower corner. The true model is given as a black line in the 1D plots and a black star for the 2D plots.

The numerical experiment shows that the EnRML can successfully approximate the true posterior solution for the GAN-FDNN modeling sequence. The numerical results show a convincing similarity between the exact samples from the posterior, acquired by the MCMC, and the approximate samples, acquired by EnRML. The EnRML provides good approximations of both the posterior earth model and the posterior input vector. Moreover, from inspection of selected elements from $\param$ it is clear that the posterior distribution can be well approximated by a Gaussian. 

\subsection{Example 2 -- Prediction of a sand-channel sequence}
\label{sec:exampleChannel}

\begin{figure}
    \centering
    a. \includegraphics[width=0.7\textwidth]{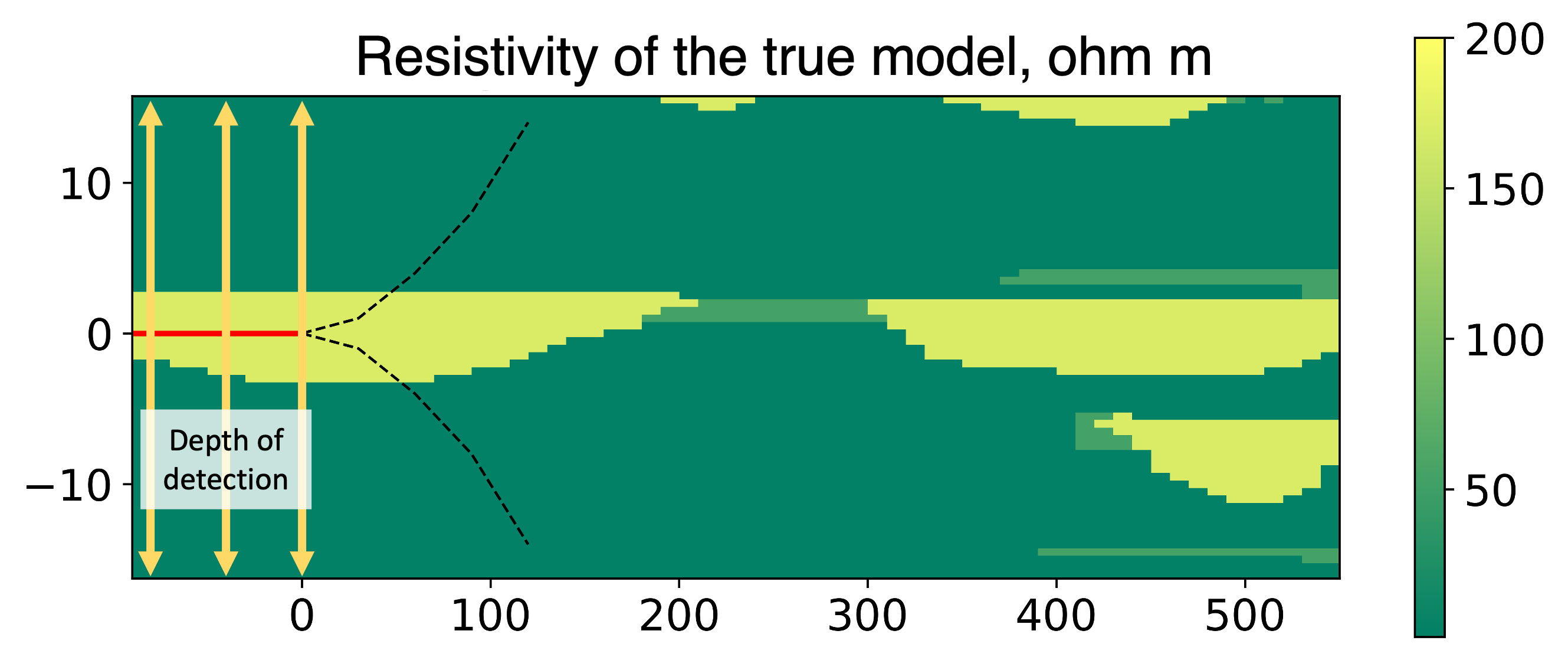}\\
    b. \includegraphics[width=0.5\textwidth]{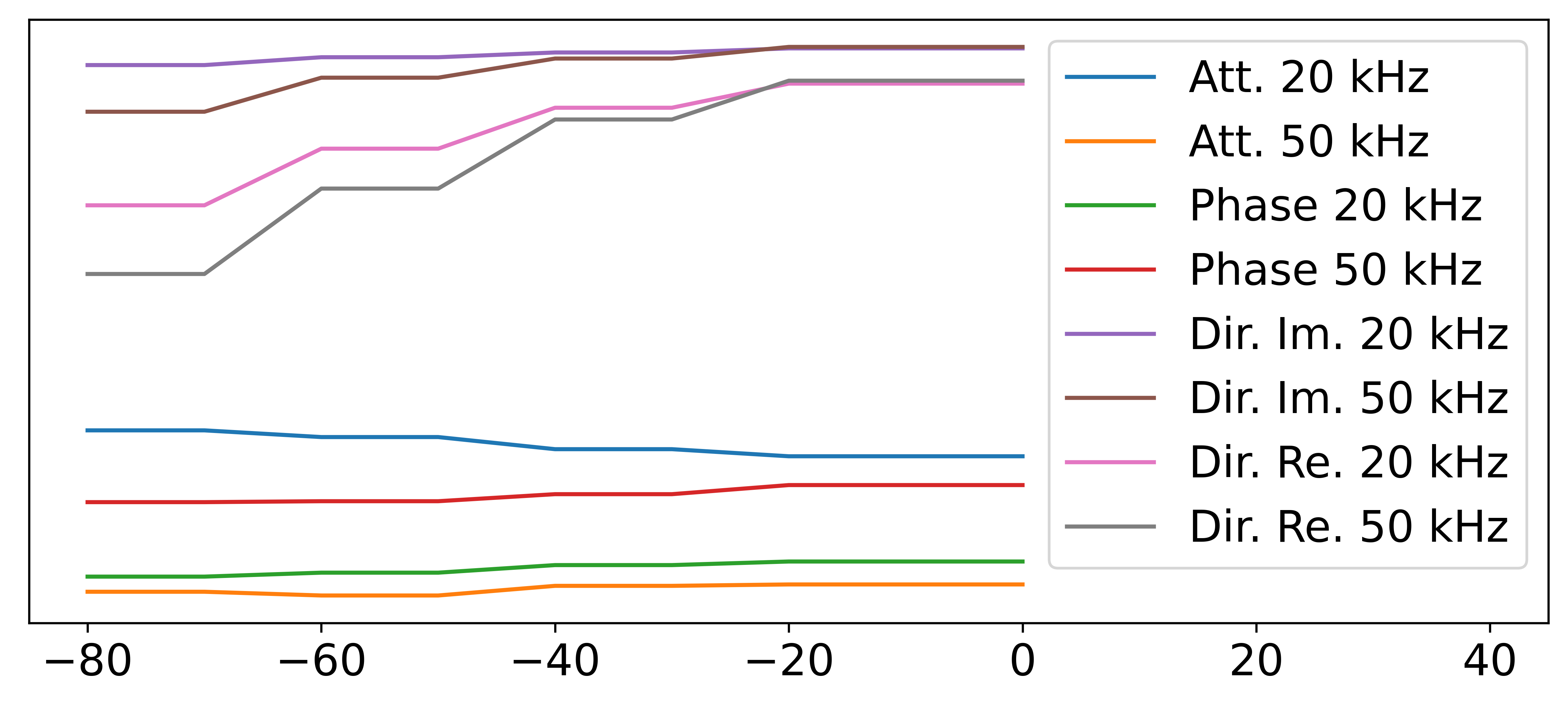}
    \caption{a. The resistivity of an earth model generated by GAN used as the synthetic truth for Example 2. 
    The yellow arrows show the region with measurements and their extent illustrate the maximum sensitivity range, termed depth of detection. 
    The filled red line is the drilled well, and the dashed lines indicate the potential for geosteering.
    b. Measurements in the eight extra-deep geophysical EM logs from the highlighted region (scaled to 0..1). The other five shallow logs not shown.}
    \label{fig:joscTruth}
\end{figure}

The second numerical example tests the ability of the workflow to predict the targets ahead of measurements in the case where the well is already landed into a sand channel, and when the pre-drill information, embedded in the prior model, is biased toward a wrong solution.
The synthetic truth for this example with the depth of detection is shown in Figure \ref{fig:joscTruth}.

 \begin{figure}
 \hspace{0.9cm} {\bf{EnRML}} \hspace{4.5cm} {\bf{MCMC}}\\
    \centering
    \includegraphics[trim=0 0 2.5cm 0, clip,height=0.22\textwidth]{fig/example2-enrml/resisitivity_mean_posterior.png}
    \includegraphics[height=0.22\textwidth]{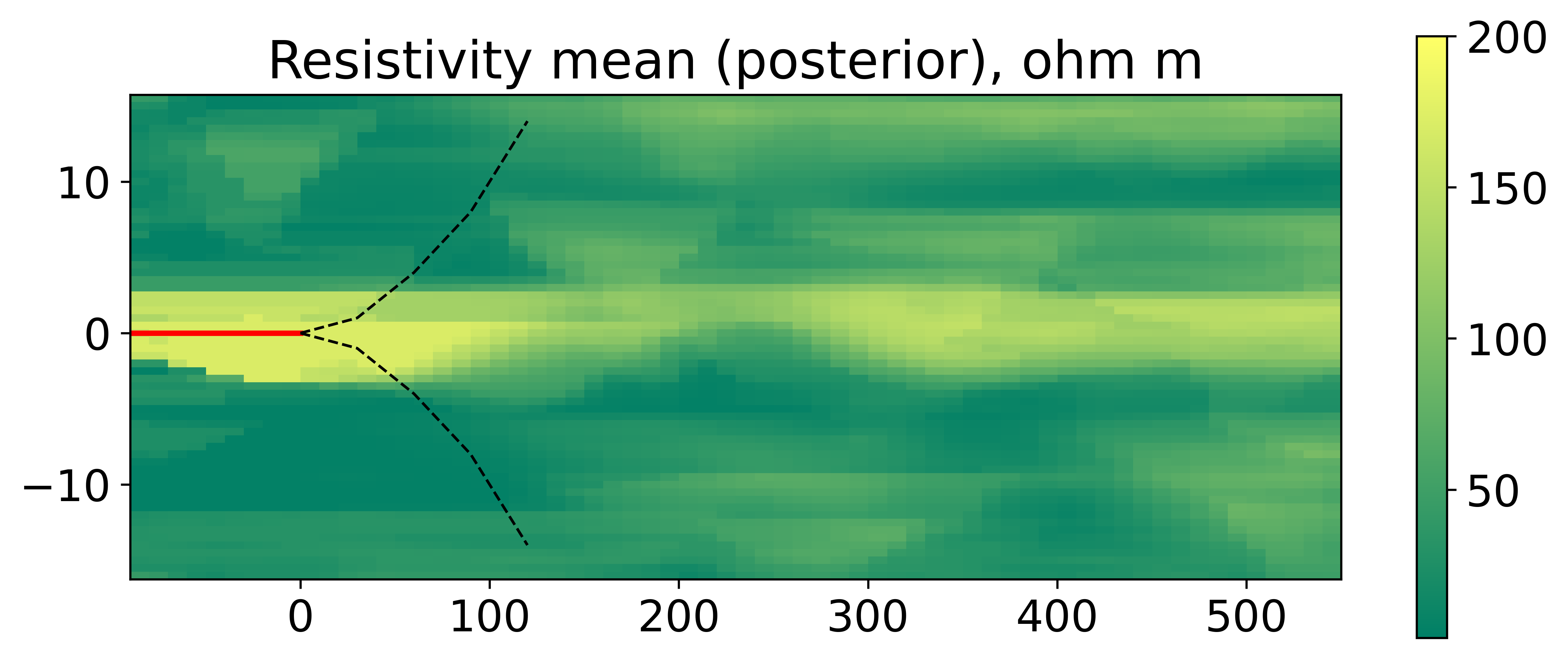}
    \\
    \includegraphics[trim=0 0 2.5cm 0, clip,height=0.22\textwidth]{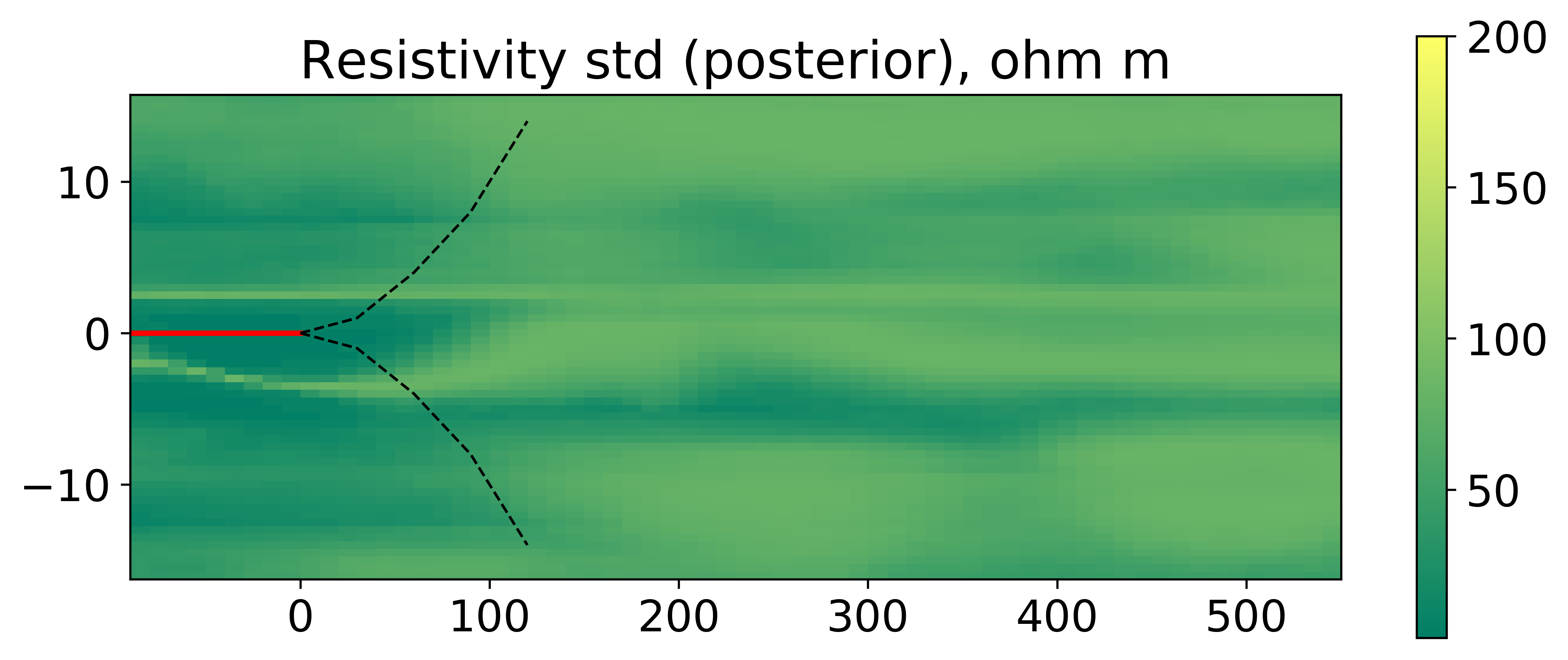}
    \includegraphics[height=0.22\textwidth]{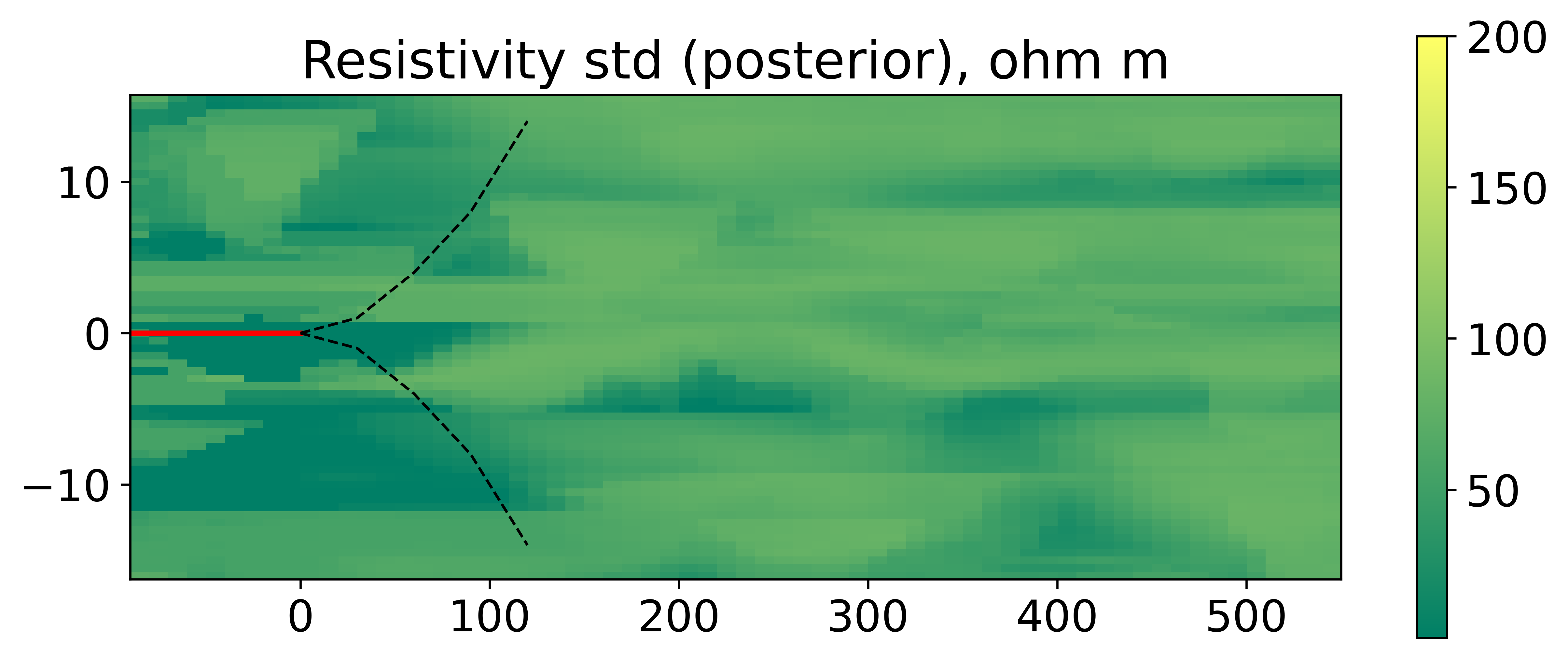}
    \caption{Mean (top) and standard deviation (bottom) in resistivity models derived from the posterior ensembles of EnRML (left column) and MCMC (right column).}
    \label{fig:resitivityPosteriorBoth-ex2}
\end{figure}
Figure~\ref{fig:resitivityPosteriorBoth-ex2} illustrates the posterior mean and standard deviation of the resistivity model. It is clear that conditioning to measurements resolves the sand channel ahead of the well position. Moreover, the EnRML does a reasonably good job in approximating the true posterior, despite the prior being slightly misspecified.

 \begin{figure}
 \hspace{0.9cm} {\bf{EnRML}} \hspace{4.5cm} {\bf{MCMC}}\\
    \centering
    \includegraphics[trim=0 0 2.5cm 0, clip,height=0.22\textwidth]{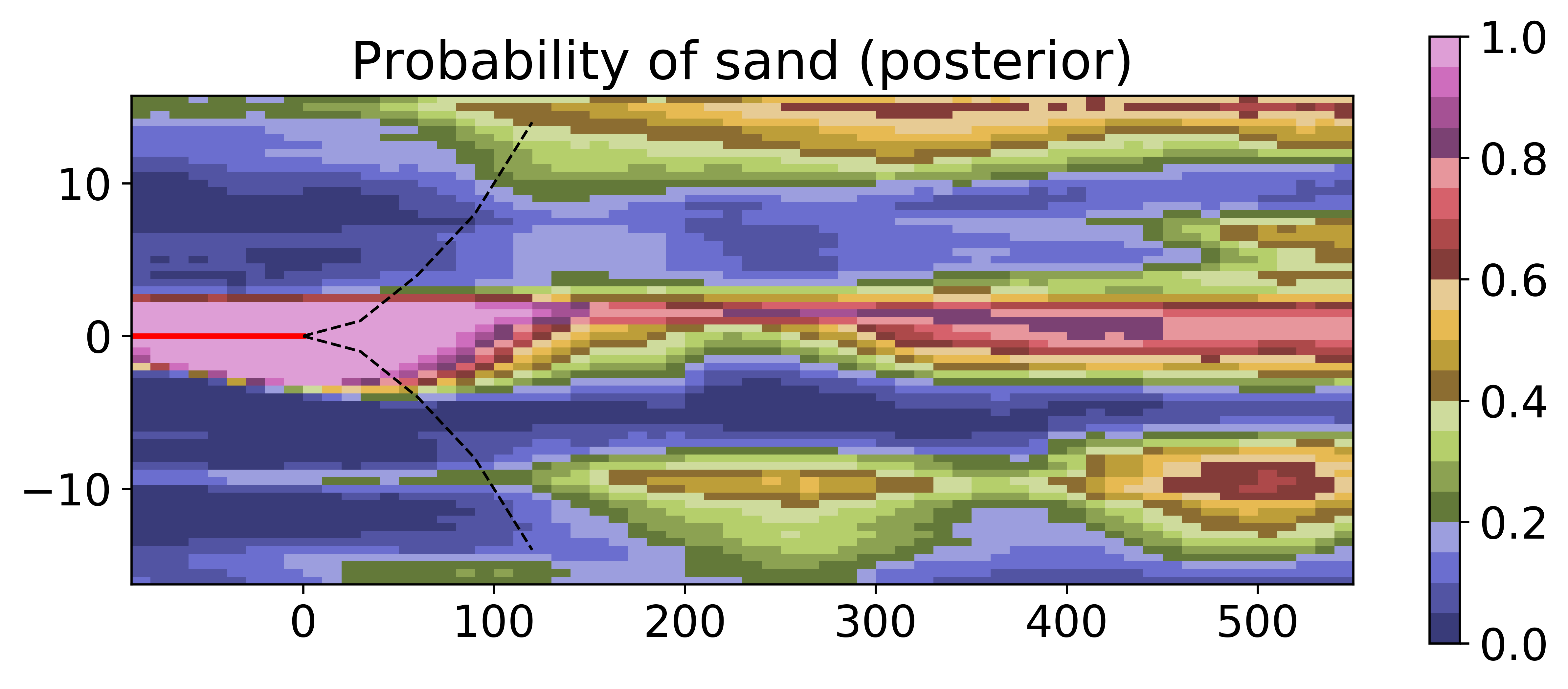}
    \includegraphics[height=0.22\textwidth]{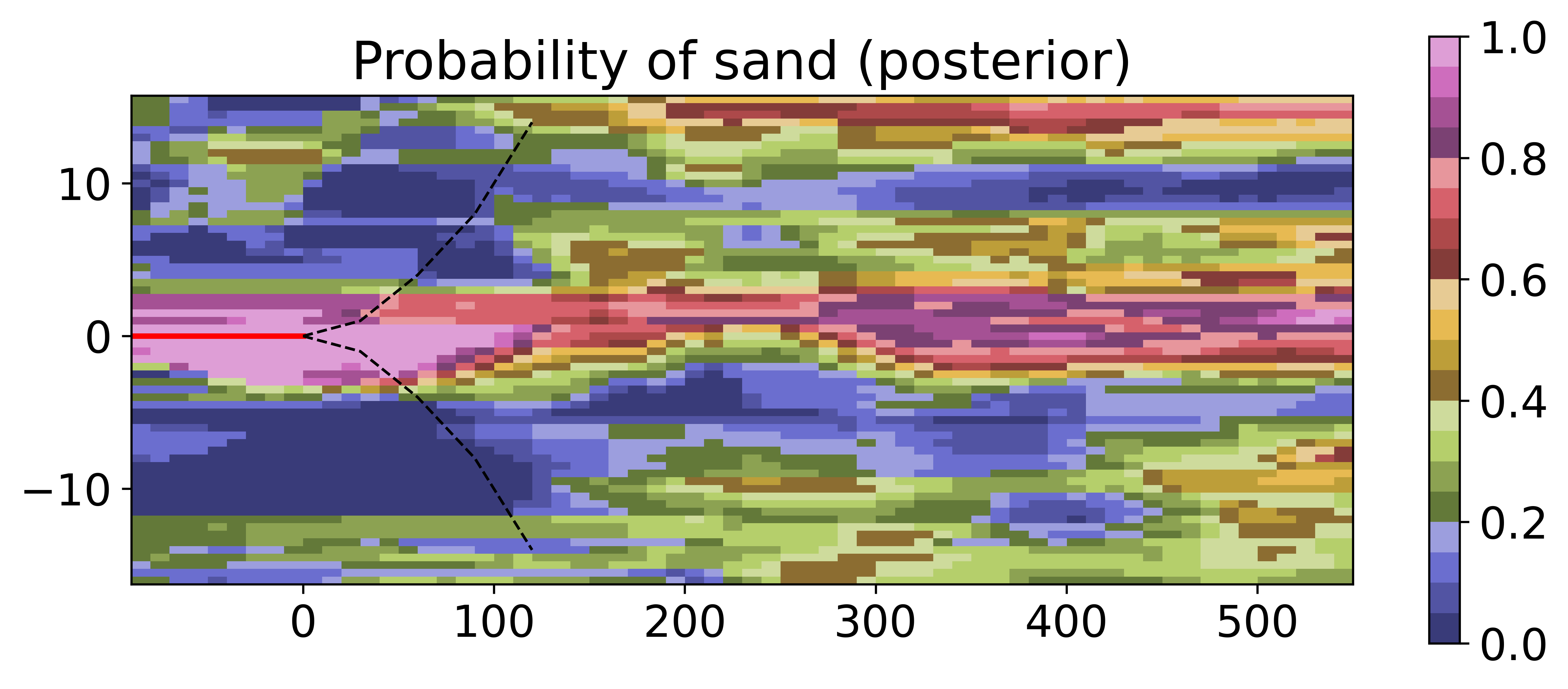}
    \caption{Probability of sand facies (crevasse or channel) from the posterior ensembles of EnRML (left column) and MCMC (right column).}
    \label{fig:prob_sand-ex2}
\end{figure}
Similarly, the estimated point-wise probability of the sand facies, plotted in Figure~\ref{fig:prob_sand-ex2}, shows that the EnRML provides excellent predictive capabilities as the sand facies is correctly forecasted to the right of the geomodel, more than 500 meters ahead of the bit. There are slightly larger differences between the EnRML and MCMC in this example. However, the approximate posterior is still very close to the MCMC posterior, especially around the well.

\begin{figure}
    \centering
    \includegraphics[width=0.7\textwidth]{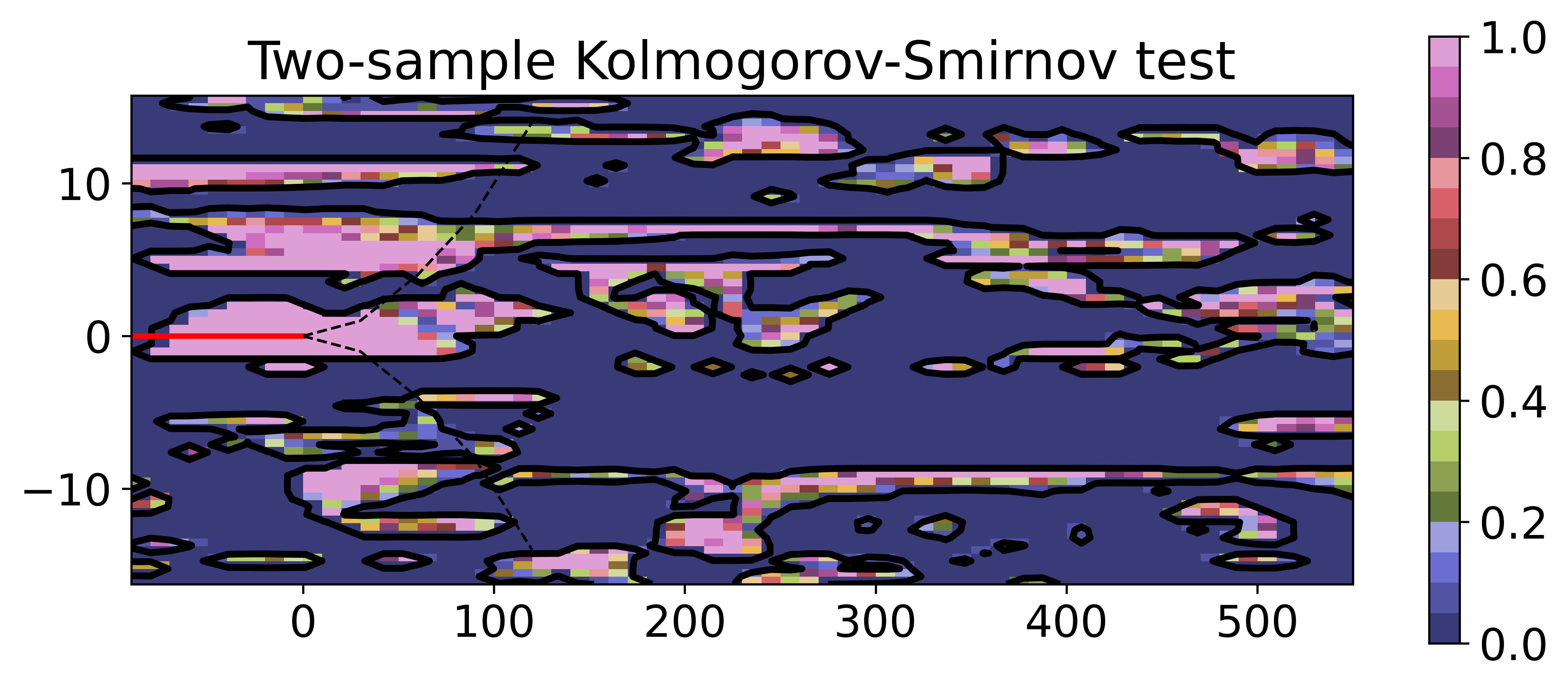}
    \caption{P-value from the Kolmogorov-Smirnov two-sample test of sand facies (crevasse or channel). The Contour line gives the significance level of 0.05.}
    \label{fig:KS_test-ex2}
\end{figure}
The measure of statistical distance between the samples from the EnRML and the samples from MCMC indicates similar performance. The P-values from the test of the H0 hypothesis of equal distributions are shown in Figure~\ref{fig:KS_test-ex2}. The significance level of 0.05 is given by the black contour line. Hence, for all p-values higher than 0.05 one cannot differentiate between the marginal distributions and the H0 hypothesis hold. Compared to example 1, there are more areas where the H0 hypothesis fails. This demonstrates that, for the more challenging experiment, there is a larger statistical distance between the approximate posterior from the EnRML and the true posterior.

\begin{figure}
    \centering
    \includegraphics[width=0.9\textwidth]{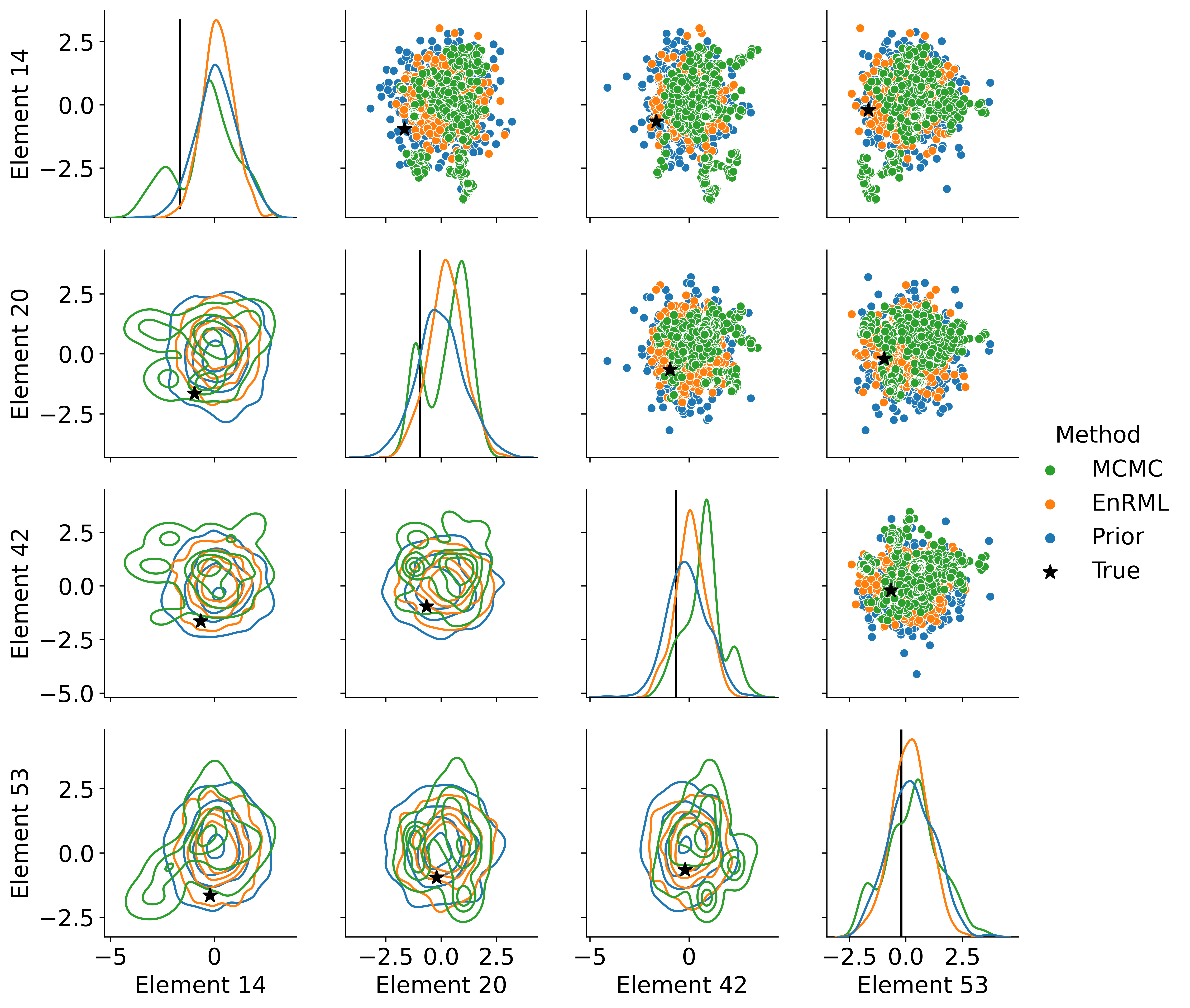}
    \caption{Marginal and bi-variate statistics of selected elements from $\param$.}
    \label{fig:Seaborn-ex2}
\end{figure}
As a final evaluation of the results, we plot a selection of marginal and bi-variate elements of the input vector $\param$. Similar to experiment 1, we highlight the effect of the biased pre-drill information by selecting two elements that were shifted ($m_{14}$ and $m_{53}$) and two that were not shifted ($m_{20}$ and $m_{342}$). In Figure~\ref{fig:Seaborn-ex2} the kernel density estimate of the selected elements is plotted along the diagonal, the scatter plot of the pairwise elements is given in the top corner, while the contours of the 2D Kernel density estimate of the pairwise elements are given in the lower corner. The true model is given as a black line in the 1D plots and a black star for the 2D plots. The effect of the biased prior can be observed in the MCMC results, where the marginal posterior is bi-modal. 

The numerical experiment shows that the EnRML can successfully approximate the true posterior solution for the GAN-FDNN modeling sequence, even when the prior model is slightly biased. The numerical results show convincing similarity between the exact samples from the posterior, acquired by the MCMC, and the approximate samples, acquired by EnRML. There is however a larger discrepancy than was observed in example 1. From inspection of selected elements from $\param$ we can observe that the biased prior results in a more non-Gaussian posterior distribution. Despite this, we claim that the EnRML provides good approximations of both the posterior earth model and the posterior input vector.

\section{Conclusions}
\label{sec:sum_conc}

In this paper, we have demonstrated that two essential parts, the earth model and the simulated extra-deep EM logs, of an ensemble-based DSS system can be substituted with neural networks. 
For the earth model, we utilize the GAN trained with images from a realistic geological setting, for the simulated logs we use a forward deep neural network (FDNN) trained using a large set of simulations from a commercial tool. 
The setup redistributes the computational cost from online to offline calculations, enabling complex earth models and deep-sensing EM logs to be part of real-time ensemble updates.  

The numerical results illustrate that the GAN-FDNN modeling sequence provides excellent probabilistic predictions ahead of drilling capturing both continuous and discrete features when conditioning to only measurements with sideways sensitivity. 
Moreover, the numerical results show that the computationally efficient EnRML algorithm can sample the true Bayesian posterior confirmed by the MCMC algorithm. 
This conclusion is valid even when the prior model is slightly biased towards a wrong solution.

The proposed approach has many beneficial factors. 
Firstly, a GAN provides large flexibility for defining the geological setting. 
Here, we consider three different facies, but one can easily imagine including features like faults and pinch-outs as well as smoothly-varying properties. 
Secondly, we only need to condition a
few parameters with Gaussian distribution
to the measurements,
which is very beneficial for the ensemble-based DA approach. 
Thirdly, since we are utilizing a neural network model to generate the simulated log, the computational cost of simulating a single ensemble member is milliseconds. Hence, the proposed approach can utilize a large ensemble for the DA part.

The numerical experiments illustrated that the posterior
has a predictive capability for both MCMC and the faster EnRML method.
The future work is to integrate the DA developed in this paper with the decision framework developed in~\cite{Alyaev2019a}, allowing DSS under a much more complex geological setting. 
Furthermore, the method can be extended to account for model errors present in machine learning approximations in real-time \cite{rammay2022probabilistic}.

\section*{Acknowledgments}
This work is part of the Center for Research-based Innovation DigiWells: Digital Well Center for Value Creation, Competitiveness and Minimum Environmental Footprint (NFR SFI project no. 309589, https://DigiWells.no). The center is a cooperation of NORCE Norwegian Research Centre, the University of Stavanger, the Norwegian University of Science and Technology (NTNU), and the University of Bergen. It is funded by Aker BP, ConocoPhillips, Equinor, Lundin Energy, TotalEnergies, Vår Energi, Wintershall Dea, Kongsberg Digital, Odfjell Drilling, Sekal, and the Research Council of Norway.

Part of the work was performed within the project ’Geosteering for IOR’ (NFR-Petromaks2 project no. 268122) which is
funded by the Research Council of Norway, Aker BP, Equinor,
Vår Energi and Baker Hughes Norway.

We would like to thank Emerson Roxar for providing an
academic licence for RMS 11.1. used for the geo-modelling in
this study.

\bibstyle{elsarticle-num}
\bibliography{Geosteering_GAN_Kristian}


\end{document}